  \documentclass[final,5p,times,twocolumn]{elsarticle}

 \usepackage{graphicx}

\usepackage{amssymb}
\usepackage{amsmath}
\usepackage{upgreek}
\usepackage{multicol}

\usepackage{lineno}

\journal{Nuclear Instruments and Methods in Physics Research A}

\begin{document}

\begin{frontmatter}



\title{Properties of a radiation-induced charge multiplication region in epitaxial silicon diodes\tnoteref{RD50frame}}
\tnotetext[RD50frame]{Work performed in the frame of the CERN-RD50 collaboration.}

 \author[hh]{J. Lange\corref{cor1}}
\ead{joern.lange@desy.de}
\author[hh]{J. Becker}
\author[hh]{E. Fretwurst}
\author[hh]{R. Klanner}
\author[hh]{G. Lindstr\"{o}m}
\cortext[cor1]{Corresponding author. Tel.: +49 40 8998 4725; fax: +49 40 8998 2959.}
\address[hh]{Institute for Experimental Physics, University of Hamburg, Luruper Chaussee 149, 22761 Hamburg, Germany}

\begin{abstract}
Charge multiplication (CM) in p$^+$n epitaxial silicon pad diodes of 75, 100 and 150~$\upmu$m thickness at high voltages after proton irradiation with 1~MeV neutron equivalent fluences in the order of $10^{16}$~cm$^{-2}$ was studied as an option to overcome the strong trapping of charge carriers in the innermost tracking region of future Super-LHC detectors. Charge collection efficiency (CCE) measurements using the Transient Current Technique (TCT) with radiation of different penetration (670, 830, 1060~nm laser light and $\alpha$-particles with optional absorbers) were used to locate the CM region close to the p$^+$-implantation. The dependence of CM on material, thickness of the epitaxial layer, annealing and temperature was studied. The collected charge in the CM regime was found to be proportional to the deposited charge, uniform over the diode area and stable over a period of several days. Randomly occurring micro discharges at high voltages turned out to be the largest challenge for operation of the diodes in the CM regime. Although at high voltages an increase of the TCT baseline noise was observed, the signal-to-noise ratio was found to improve due to CM for laser light. Possible effects on the charge spectra measured with laser light due to statistical fluctuations in the CM process were not observed. In contrast, the relative width of the spectra increased in the case of $\alpha$-particles, probably due to varying charge deposited in the CM region.

\end{abstract}


\end{frontmatter}


\section{Introduction}
\label{intro}
After approximately 10 years of operation, an upgrade of the LHC to a tenfold luminosity of 10$^{35}$~cm$^{-2}$s$^{-1}$ (Super-LHC or SLHC) is under discussion. This will lead to very high radiation levels, especially in the innermost tracking region where a 1~MeV neutron equivalent fluence of $\Phi_{eq}=1.6\times 10^{16}\mbox{ cm}^{-2}$ is expected \cite{Gia02}. Thus silicon detectors as used today will suffer from severe damage caused by radiation-induced defects. A strong degradation of the signal-to-noise ratio (SNR) will occur due to an increasing depletion voltage ($U_{dep}$) at high fluences, higher leakage currents ($I_{rev}$) and trapping of charge carriers.

\begin{figure}[bt]
	\centering
		\includegraphics[width=8.5cm]{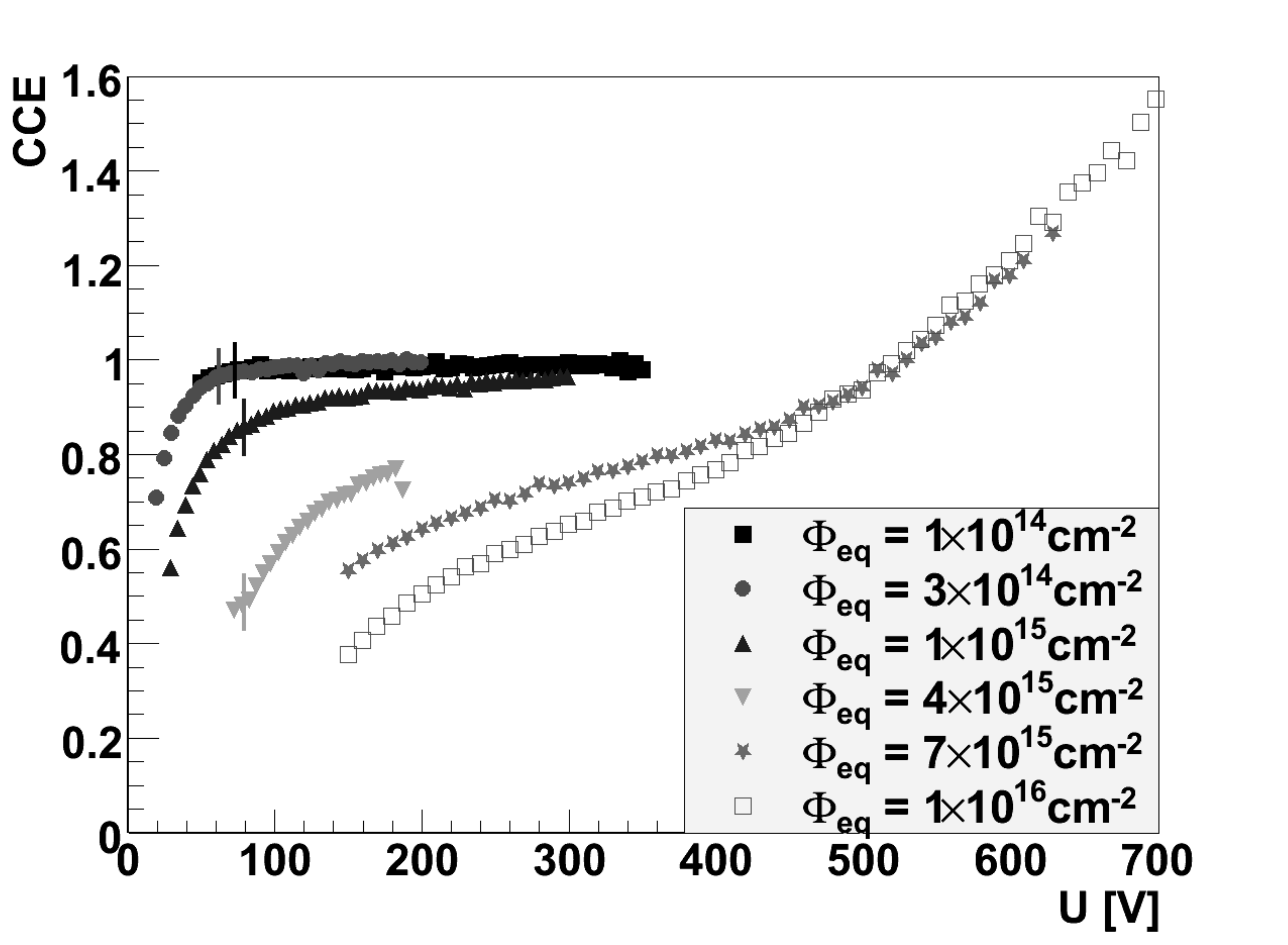}
	\caption{The CCE in n-type EPI-ST 75~$\upmu$m after proton irradiation measured with $\alpha$-particles as a function of bias voltage for different fluences. Measurements up to $4\times10^{15}$~cm$^{-2}$ were done at room temperature after 60~min annealing at 80$^{\circ}$C. The two highest fluences were measured at -10$^{\circ}$C after 30~min annealing at 80$^{\circ}$C. The vertical lines indicate $U_{dep}$ taken from capacitance-voltage (CV) measurements at room temperature and 10~kHz.}
	\label{CCE-intro}
\end{figure}

Trapping reduces the charge collection efficiency (CCE) significantly and is considered to be the most limiting factor at SLHC fluences. The degradation of CCE at moderate voltages (e.g. around 150~V) with increasing fluence is clearly visible in Fig.~\ref{CCE-intro}. However, recently we reported on CCE values larger than 1 in highly proton-irradiated thin epitaxial (EPI) diodes at high voltages (see Fig.~\ref{CCE-intro}), which indicates that trapping is overcompensated by charge multiplication (CM) \cite{Lan08, Lan10}. Apparently, the electric fields in thin, highly-irradiated sensors operated at high voltages are sufficiently high for impact ionisation. Evidence of CM has been also found in CCE measurements of highly-irradiated planar \cite{Man09, Kra09, Mik10, Cas09} and 3D \cite{Koe10} strip detectors. Obviously, it would be desirable to use CM effects to overcome trapping in highly-damaged position-sensitive detectors at the SLHC. Therefore, a detailed understanding of the formation and the properties of such a radiation-induced CM region is of significant interest.

Consequently, the following issues related to CM are studied in this work: the location of the CM region with respect to the detector depth, the dependence of CM on thickness, material, annealing and temperature, the proportionality between measured and deposited charge, the spatial uniformity of CM over the detector area, the long-term stability of detector operation in the CM regime and effects of CM on the charge spectrum and noise.

\section{Investigated material, samples and irradiation}
Epitaxial p$^+$n pad diodes of various oxygen content and thicknesses were studied. Epitaxial silicon is considered a promising candidate for superior radiation hardness because it is a combined approach of \emph{defect engineering} (high oxygen concentration, which can be further enhanced by oxygenation) and \emph{device engineering} (thin layers: 25-150~$\upmu$m) \cite{Lin06}. In the case of proton irradiation, such material is characterised by predominant donor introduction so that no space charge sign inversion (SCSI) occurs for n-type bulk doping \cite{Pin09}. Moreover, the average electric field is higher in thin sensitive layers, which will play a crucial role for this work.

N-type epitaxial layers of 75~$\upmu$m ($\approx$150~$\Upomega$cm resistivity), 100~$\upmu$m ($\approx$300~$\Upomega$cm) and 150~$\upmu$m ($\approx$500~$\Upomega$cm) thickness were grown on a highly-doped $\approx$500~$\upmu$m thick Cz substrate by ITME \cite{ITME}. Both standard (\emph{ST}) and oxygen-enriched (\emph{DO}) materials were studied. ST material showed a nonuniform oxygen distribution with average concentrations of (5-9)$\times10^{16}$~cm$^{-3}$, increasing with decreasing thickness. The oxygen distribution of DO material was more uniform with average concentrations between $1.4\times10^{17}$~cm$^{-3}$ (150~$\upmu$m) and $6\times10^{17}$~cm$^{-3}$ (75~$\upmu$m). Pad diodes of $5\times5$~mm$^2$ (labelled as \emph{big}) and $2.5\times2.5$~mm$^2$ (\emph{small}) active area with a surrounding guard ring were processed at CiS \cite{CiS}. The metallisation of the front side, where the p$^{+}$n junction is located, has an optical window for illumination.

Irradiations with 24 GeV/c protons up to equivalent fluences of $10^{16}$~cm$^{-2}$ were performed at the CERN PS irradiation facility \cite{PIrS} with an average proton flux of several 10$^{9}$~cm$^{-2}$s$^{-1}$ at a temperature between 27 and 29$^\circ$C. Proton fluences were measured by Al foil activation with a quoted accuracy of 10\% and scaled to 1~MeV neutron equivalent fluences, which will be referred to in the following, using a hardness factor of $\kappa=0.62$.

If not stated otherwise, the samples were annealed for 30~min at 80$^{\circ}$C.

Most of the following investigations will focus on a big diode of EPI-ST~75~$\upmu$m material irradiated by $10^{16}$~cm$^{-2}$ (referred to as \emph{75-1E16}) as CM turned out to be very pronounced in this sample with the highest fluence and the smallest thickness.

\section{Experimental methods}
\label{Sec:methods}

\subsection{TCT and charge collection}
Charge collection was investigated with the Transient Current Technique (TCT). Different radiation was used to produce free charge carriers that drift under the influence of an applied reverse bias voltage towards the n$^+$- (electrons) or the p$^+$-electrode (holes). Each drifting electron (e) or hole (h) induces a current given by

\begin{equation}
\label{Ramo}
I_{e,h}(t) = \frac{q_0}{d} v_{dr_{e,h}}(E(x(t))) ,
\end{equation}
where $q_0$ is the elementary charge, $1/d$ the inverse sensor thickness, i.e. the weighting field in pad diodes, and $v_{dr_{e,h}}$ the drift velocity that depends on the electric field $E$ \cite{Ram39}. The total transient current is the sum of all individual currents and therefore depends on the number of charge carriers $N$ and their distribution inside the detector. $N$ might decrease or increase with time due to trapping or CM. The trapping probability, i.e. the inverse effective trapping time $1/\tau_{eff}$, was found to be proportional to the fluence \cite{Kra02}. However recently, in irradiated EPI material observations have been made that $1/\tau_{eff}$ is less at high voltages respectively fields than predicted \cite{Lan10}. CM can be described by ionisation coefficients $\alpha_{n,p}$, which give the ionisation probability per unit distance and are a strong function of electric field \cite{Gra73}. An important parameter is the ratio between hole and electron ionisation, $k=\alpha_{p}/\alpha_{n}$, which is much less than 1 for low fields in Si, but increasing with $E$.

The collected charge $Q$ was determined as the integral of the current pulse and the CCE was then obtained by normalising $Q$ to $Q_0$ of an unirradiated diode. It should be noted that the CCE is influenced by both trapping and CM so that it is difficult to disentangle these two effects.

\subsection{Types of radiation}
\begin{figure}[bt]
	\centering
		\includegraphics[width=8.5cm]{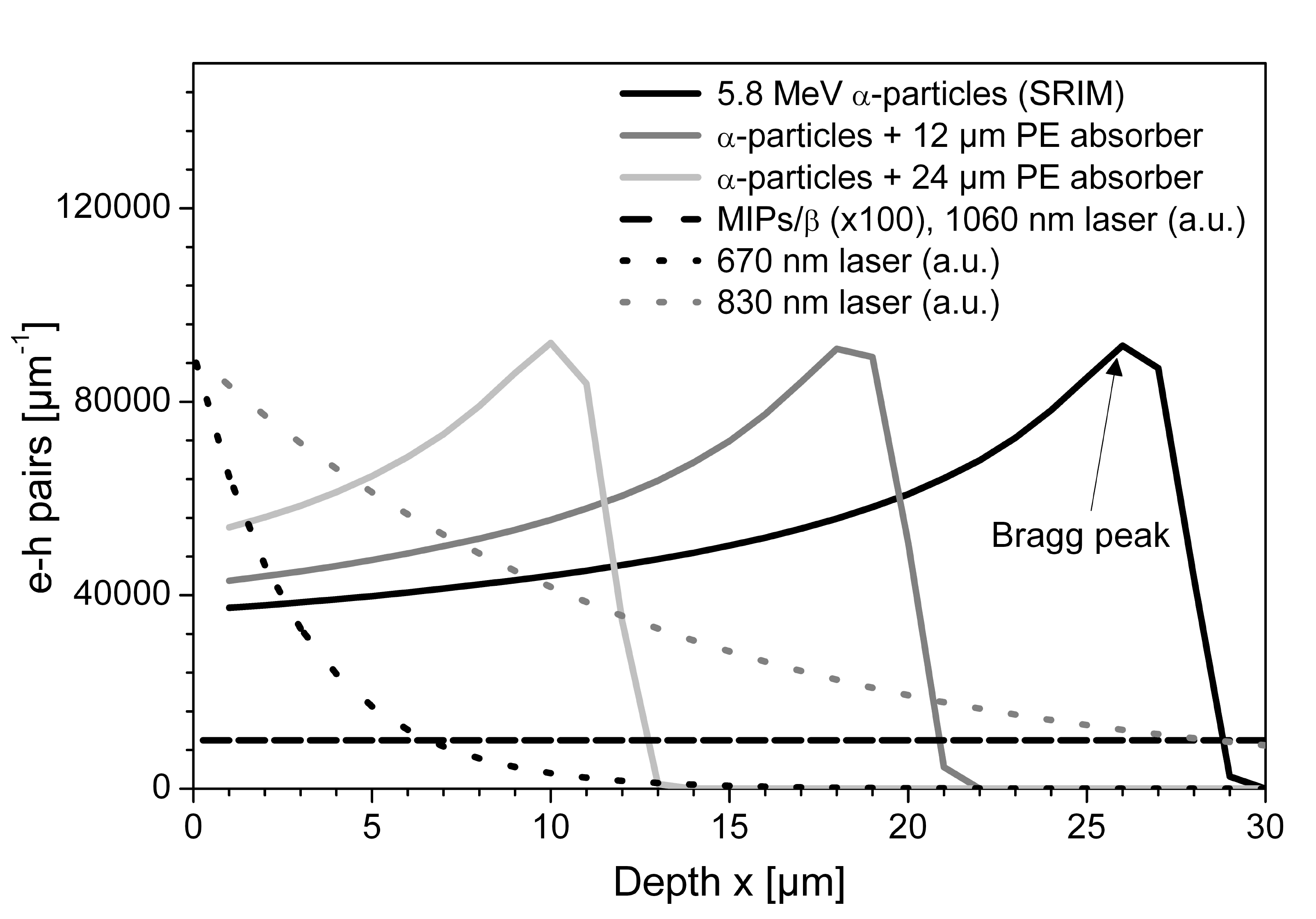}
	\caption{The distribution of electron-hole pairs deposited by different radiation as a function of detector depth. The Bragg curves for $\alpha$-particles were simulated using SRIM \cite{SRIM} taking into account absorption in air, PE and the dead layers of SiO$_2$ and the p$^+$-implant (perpendicular incidence). The scale for laser light is arbitrarily normalised.}
	\label{penetration}
\end{figure}

In this work radiation with different penetration was used. For epitaxial diodes only front illumination is possible due to the thick insensitive Cz substrate at the back side. The initial distributions of deposited electron-hole (e-h) pairs as a function of distance from the p$^+$n junction are shown in Fig.~\ref{penetration}.

Red laser light of 660 or 670~nm wavelength ($\approx$3~$\upmu$m absorption length) produces charge carriers close to the illuminated front surface so that the current signal is almost entirely due to electrons in n-type material. In contrast, 1060~nm laser light deposits charge almost uniformly over the whole detector thickness similarly to minimum ionising particles (MIPs) so that the current is due to both electrons and holes. The penetration of 830~nm laser light with an absorption length of $\approx$13~$\upmu$m lies in between. 5.8~MeV $\alpha$-particles from a $^{244}$Cm source feature a Bragg peak around 26~$\upmu$m deep inside the detector for perpendicular incidence. Polyethylene (PE) absorbers of different thicknesses (12 and 24~$\upmu$m) can be inserted between the source and the sample in order to deliberately modify the penetration and shift the Bragg peak closer to the surface. However, there is also unintended reduction of effective penetration depth for part of the beam due to $\approx$50$^\circ$ divergence allowed by the collimator and single $\alpha$-particles with lower energy ($\approx$5\% of the beam), possibly caused by energy loss at the edge of the plastic collimator.

\subsection{TCT setups}
The TCT setup for $\alpha$-particles consisted of a brass box in which the diode could be mounted and a lid with the $\alpha$-source in the middle of a plastic collimator, which was placed about 2~mm above the centre of the diode. The diode was biased at the front side with a Keithley 6517A up to 1000~V and read out over a Picosecond Pulse Lab 5531 Bias-T with a 1.8~GHz current-sensitive Phillips Scientific amplifier with a gain of 100 and a 1~GHz Tektronix DPO~4104 oscilloscope with 5~GS/s. The signal from $\alpha$-particles occurs randomly so that self-triggering with a threshold of 80~mV was needed. The achieved CCE precision was estimated to be within 3\%.

CCE measurements with 670 and 1060~nm laser light were done using lasers from Advanced Laser Systems with a pulse width below 100~ps. The laser intensity was tunable both with the laser generator and with additional optical attenuators so that (0.8 to 50)$\times 10^{6}$~e-h pairs could be produced in the diode with the 670~nm laser with $\approx$1.8$\times 10^{6}$~e-h pairs chosen as standard. For 1060~nm laser light the deposited charge depends on the diode thickness and temperature, but usually not more than a few $10^{6}$~e-h pairs were produced. An 830~nm laser diode was driven with a pulse generator and delivered light pulses of $\approx$1~ns duration which produced $\approx$4.5$\times 10^6$~e-h pairs. The light was sent via an optical fibre to the illumination window without additional focusing in order to avoid high charge carrier densities. A repetition rate of 50~Hz was chosen and the trigger for the oscilloscope was provided externally by the laser driver. The readout was either the same as for the setup with $\alpha$-particles (front-bias) or optionally back-bias without Bias-T readout could be chosen, but within the quoted uncertainties the CCE results were the same. A CCE accuracy of about 2\% was achieved.

Another setup developed for multi-channel TCT was used in order to perform x-y scans over the illumination window in the front surface. The light of a 660~nm Picoquant laser with less than 100~ps pulse width and 1~kHz repetition rate was focused to a measured spot size of $\sigma_{spot}$=20~$\upmu$m. Only 6$\times10^5$~e-h pairs were injected to avoid effects of high charge carrier densities. The sample was movable on x- and y-tables with 0.1~$\upmu$m precision. Front side readout as explained for the setup with $\alpha$-particles was applied, but as oscilloscope a Tektronix DPO~7254 with 2.5~GHz bandwidth and 40~GS/s was used.

In all cases cooling could be applied either by a chiller with a temperature precision of around 0.3$^\circ$C or by a chiller in combination with a Peltier element, which improved the precision to better than 0.1$^\circ$C. If not stated otherwise, the measurements were done at -10$^\circ$C in a nitrogen atmosphere in order to reduce the high leakage current and operate the diode under LHC conditions. Furthermore, if not stated otherwise, 512 current pulses were averaged.

Due to high capacitances and small charge collection times in the order of 1~ns in thin diodes the time structure of the signal was not well-resolved except for small 150~$\upmu$m thick sensors so that only the collected charge will be dealt with in the following.

A problem during the measurements at high voltages of both irradiated and unirradiated diodes were micro discharges, i.e. randomly occurring current peaks probably due to sudden discharges in high-field regions at the edge of the diode or implants. The occurrence and frequency were not reproducible. If self-triggering is used, micro discharges can produce signals above the trigger threshold and in case of a high discharge rate no measurements are possible anymore.

\section{Localisation of the CM region}
\label{Sec:Loc_d}

\begin{figure*}[p]
	\centering
		\includegraphics[width=13cm]{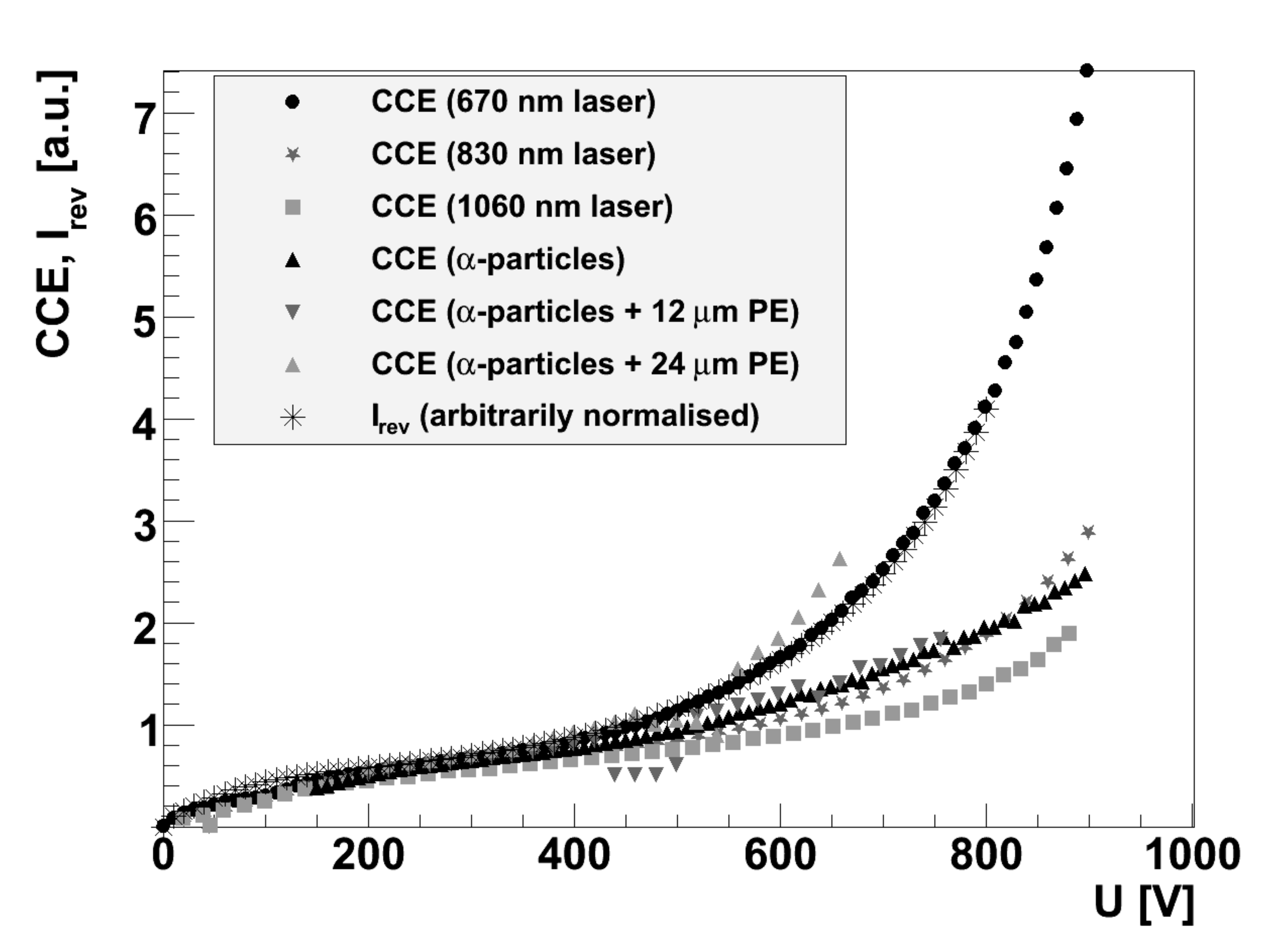}
	\caption{The CCE as a function of bias voltage in 75-1E16 measured with different types of radiation and $I_{rev}$ (arbitrarily normalised).}
	\label{CCE_different_pen}
\end{figure*}

\begin{figure*}[p]
	\centering
		\includegraphics[width=8.5cm]{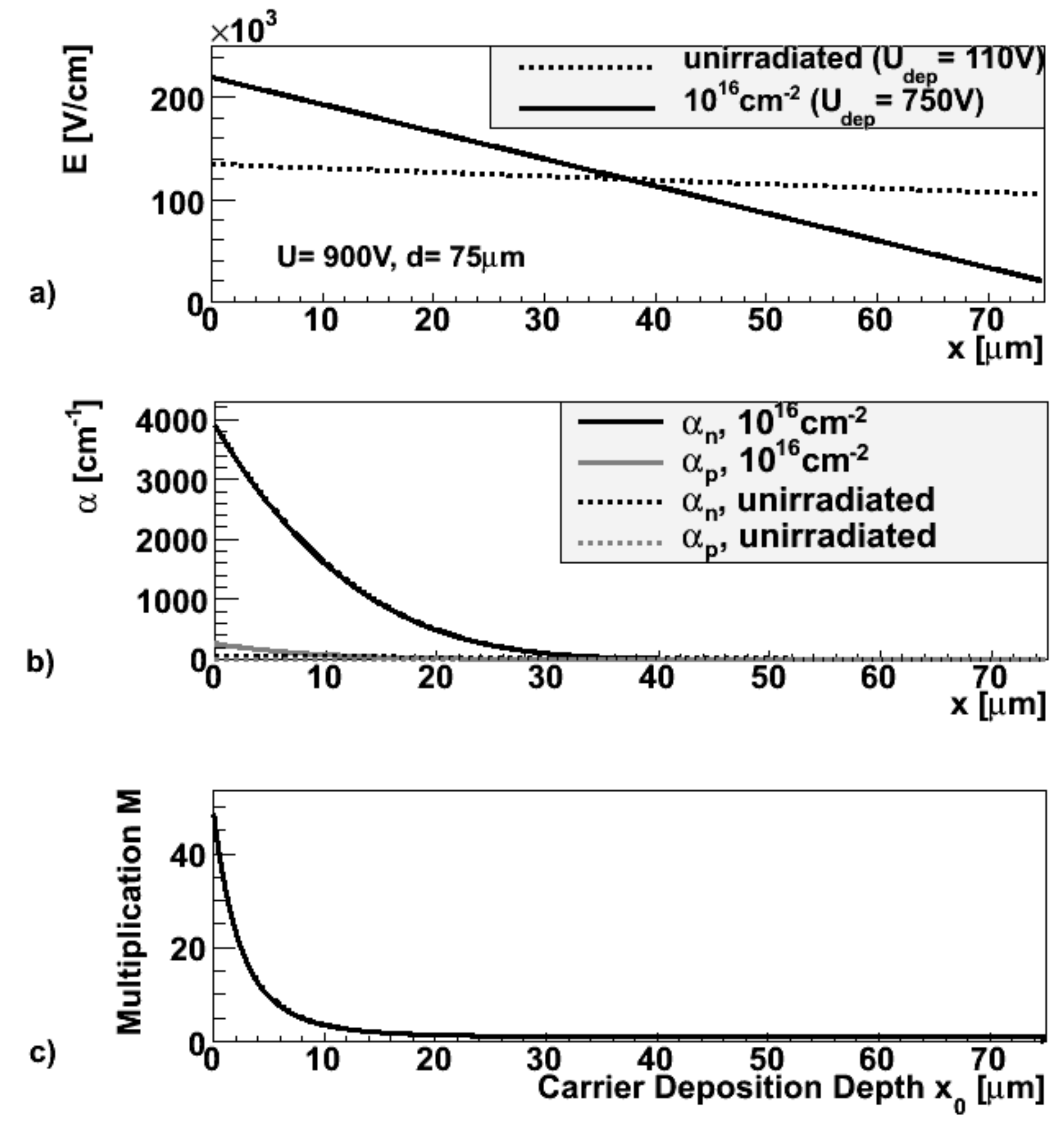}
		\includegraphics[width=8.5cm]{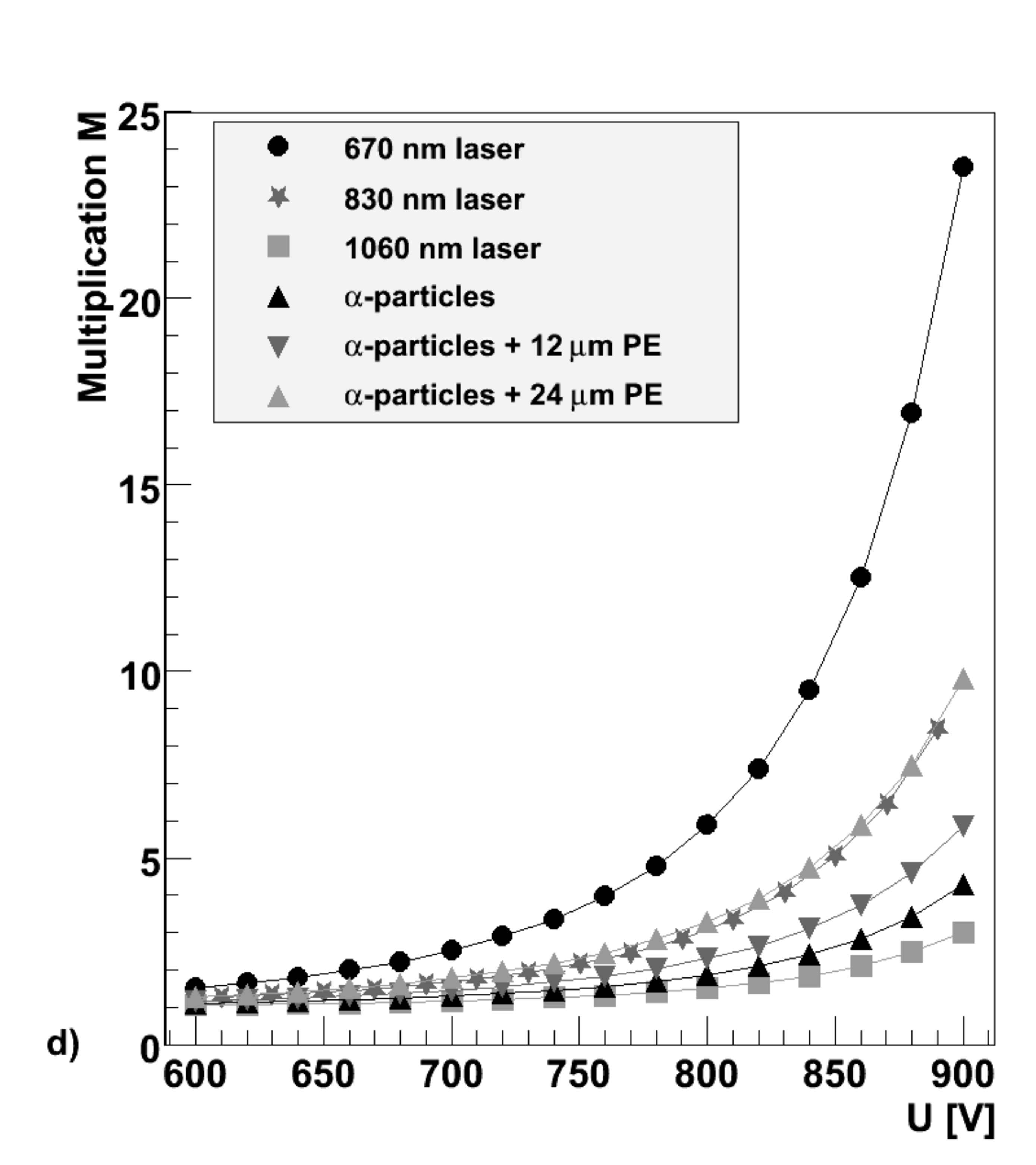}
	\caption{Qualitative explanation for the development of a radiation-induced CM region. a) Electric field distribution $E(x)$ in the linear model for an unirradiated 75~$\upmu$m diode and for 75-1e16 at 900~V. b) The corresponding distributions for the ionisation coefficients $\alpha_n(x)$, $\alpha_p(x)$ according to \cite{Gra73}. c) The resulting multiplication $M$ as a function of deposition depth $x_0$. d) $M$ for different types of radiation as a function of bias voltage.}
	\label{CM_simulation}
\end{figure*}

Radiation with different penetration was used to investigate where the CM region is located with respect to the detector depth. Results for 75-1E16 with 670 and 1060~nm laser light and $\alpha$-particles without absorber were already presented in \cite{Lan10}. Fig.~\ref{CCE_different_pen} shows them together with additional measurements using 830~nm laser light and $\alpha$-particles with 12 and 24~$\upmu$m PE absorber. The self-triggered measurements with $\alpha$-particles with 12 and 24~$\upmu$m absorber gave sometimes low values in certain voltage ranges due to suddenly occurring micro discharges. It can be seen that also the reverse current does not saturate but rises strongly due to multiplication.

At high voltages the CCE of 670~nm laser light and $\alpha$-particles with 24~$\upmu$m absorber is highest, followed by $\alpha$-particles with 12~$\upmu$m absorber and without absorber and by 830~nm laser light. The lowest CCE, but still exceeding 1, was measured for 1060~nm laser light. To conclude, the CM depends on the position where the e-h pairs have been initially deposited: The closer the charge carriers have been produced to the front surface the stronger the CM.

The following simplified considerations on the formation of the radiation-induced CM region show that this is what one would expect because the high-field region in non-type-inverted p$^+$n diodes is located at the front surface. For a first rough estimation a linear field dependence with a depletion voltage of 750~V due to uniform radiation-induced space charge is assumed. Unfortunately, $U_{dep}$ could not be directly measured for this sample due to high leakage currents, but needed to be extrapolated from measurements of the donor introduction rate in lower irradiated diodes after 8~min annealing at 80$^{\circ}$C \cite{Lan08}. Also the assumption of a linear electric field dependence is only a rough approximation in highly-irradiated diodes as the high level of reverse current modifies the electric field distribution by changing the space charge and leading to a voltage drop over the neutral bulk region, which becomes highly-resistive after irradiation \cite{Ere02, Ver07}. E.g. double peak electric field structures have been observed in 150~$\upmu$m thick EPI diodes irradiated with fluences around $10^{15}$~cm$^{-2}$ \cite{Lan10}. Fig.~\ref{CM_simulation}a shows the assumed field at 900~V compared to the one of an unirradiated diode with $U_{dep}=110$~V. The maximum of $E$ at the front side is higher for 75-1E16 than for the unirradiated case. As the ionisation coefficient $\alpha(E)$ is a strong function of the electric field, such a difference in $E$ leads to considerable values of $\alpha_n$ for electrons in 75-1E16 in a 30~$\upmu$m thick layer (Fig.~\ref{CM_simulation}b). In contrast, $\alpha_p$ for holes in 75-1E16 is much less and the $\alpha$-values in the unirradiated case are 0. The consequences of such a distribution of $\alpha_n$ for CM is shown in Fig.~\ref{CM_simulation}c where the multiplication factor (or gain) $M(x_0)=\exp(\int_{x_0}^d{\alpha_n(E(x))dx})$ is displayed (neglecting hole multiplication). $M(x_0)$ gives the number of collected charge carriers in the absence of trapping for one e-h pair that has been initially deposited at $x_0$. It can be seen that considerable values of $M$ are present in a thin, approximately 10~$\upmu$m thick layer at the front side. If $M(x_0)$ is convoluted with the respective distributions of deposited charge from Fig.~\ref{penetration}, multiplication factors for the different kinds of radiation can be calculated, which is shown in Fig.~\ref{CM_simulation}d as a function of voltage. For a realistic calculation of the CCE trapping effects have to be included. But if Fig.~\ref{CM_simulation}d is qualitatively compared to the CCE measurements in Fig.~\ref{CCE_different_pen} one can see that the order of the curves for different laser light wavelengths are well-reproduced, as well as the order of the ones for $\alpha$-particles with different absorbers. Only when cross-comparing the curves for laser light and $\alpha$-particles inconsistencies occur as the curves for $\alpha$-particles are measured to be higher than simulated compared to the laser curves. This can be explained by the components of the $\alpha$-beam with effectively shallower penetration due to its divergence or low-energy component (cf. Section~\ref{Sec:methods}) because they undergo stronger multiplication. Moreover, the calculated curves might look different if trapping is taken into account as also the effect of trapping depends on the distribution of deposited charge carriers.

As the effect of CM is the biggest for 670~nm laser light due to its shallow penetration, most of the following measurements were performed with this type of radiation.

\section{CM for different materials and thicknesses}
\label{Sec:different_material}
\begin{figure}[!h]
	\centering
		a)\includegraphics[width=8.5cm]{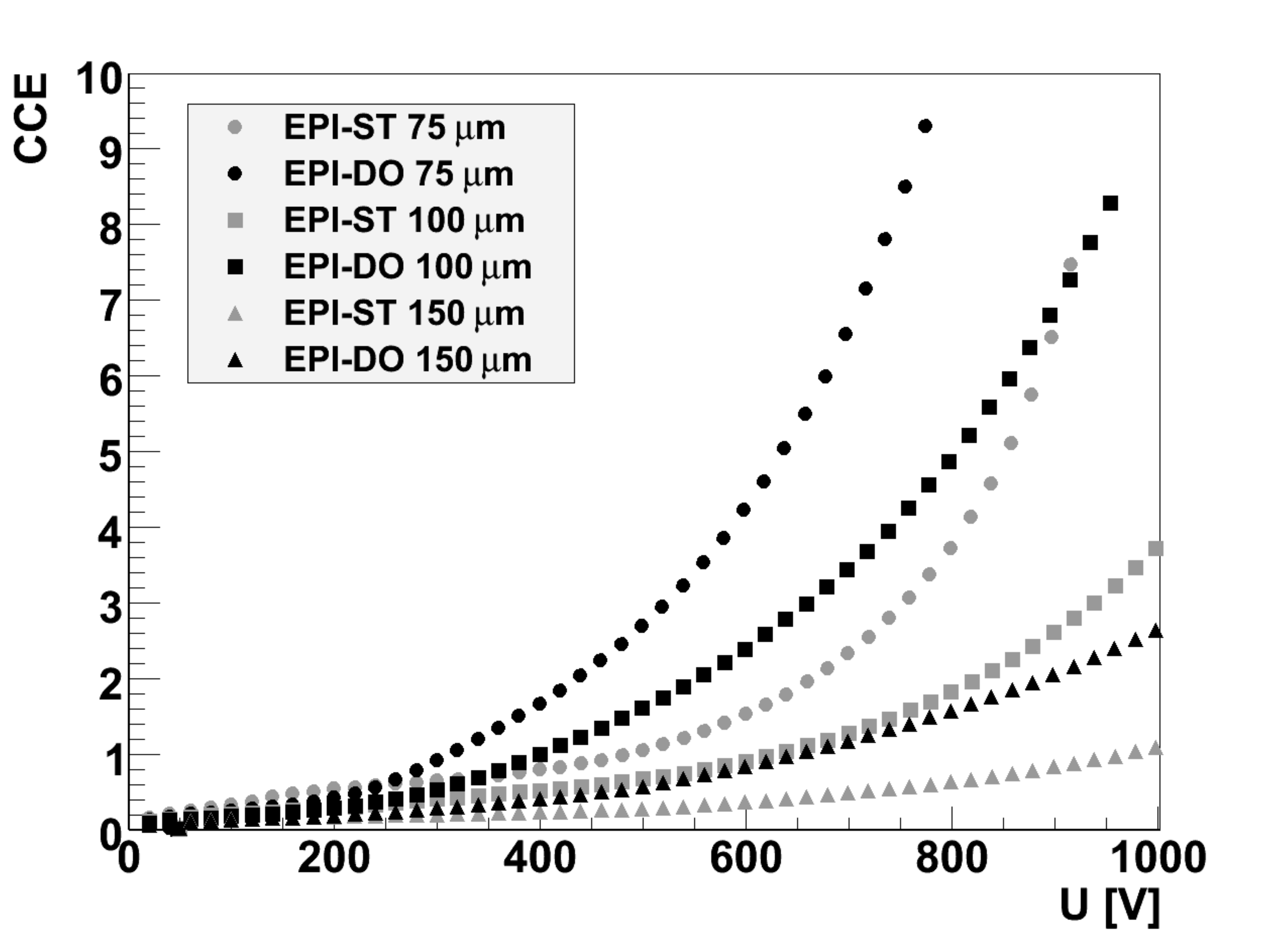}
		b)\includegraphics[width=8.5cm]{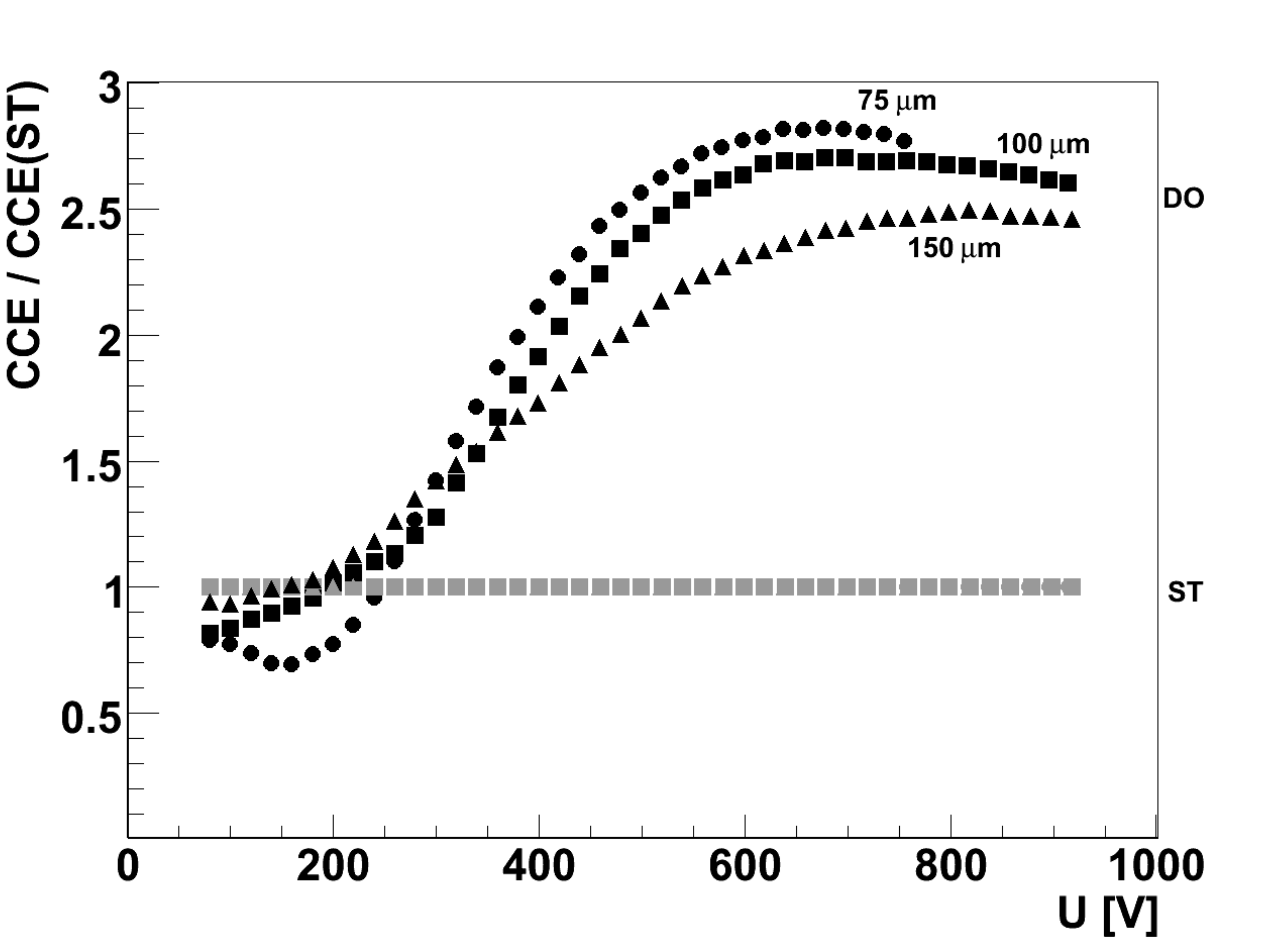}
		c)\includegraphics[width=8.5cm]{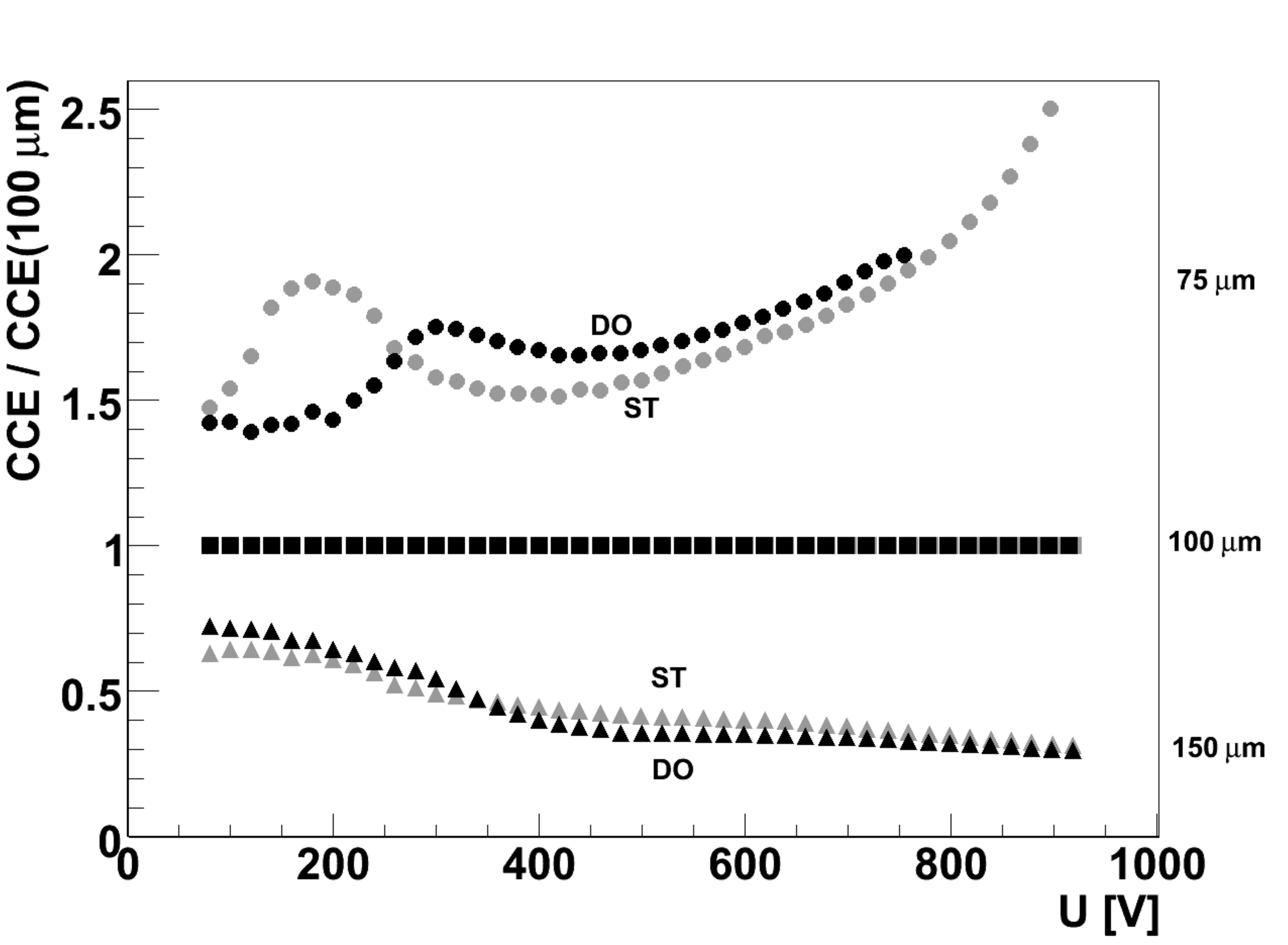}
	\caption{a) The CCE as a function of bias voltage for different materials and thicknesses at a fluence of $\Phi_{eq}=10^{16}$~cm$^{-2}$ measured with 670~nm laser light. b) The CCE normalised to the CCE of ST material with the same thickness. c) The CCE normalised to the CCE of 100~$\upmu$m for the same material.}
	\label{different_material}
\end{figure}

Fig.~\ref{different_material}a shows the CCE as a function of bias voltage for different materials (ST and DO) and thicknesses at a fluence of $10^{16}$~cm$^{-2}$ measured with 670~nm laser light. For a fluence of $7\times10^{15}$~cm$^{-2}$ the same material and thickness dependence was found.

In the CM regime the CCE of DO material is observed to be higher than the CCE of ST material with the same thickness. The ratio between the CCE of DO and ST material increases similarly for all thicknesses from values below 1 at low voltages to around 2.5 at high voltages (see Fig.~\ref{different_material}b). This can be explained by a higher donor introduction rate for DO material compared to ST material as observed in n-type EPI diodes \cite{Lan08}, which leads to higher space charge densities and therefore higher maximum field values at the p-n junction. Probably also the different distributions of oxygen concentration affect the space charge and consequently the field distribution.

The CCE of thinner diodes is found to be higher than the CCE of thicker samples for the same Si material (see Figs.~\ref{different_material}a,c). Not the whole effect can be attributed to a difference in CM, but also the strong trapping in combination with the different weighting fields $1/d$ must be taken into account. Assuming for simplicity that all charge carriers are created and multiplied with $M$ at x=0 and trapped after the mean drift distance\footnote{Assuming a drift with the saturation velocity $v_{sat,e}\approx10^7$~cm/s and a trapping time constant of $\tau_{eff,e}\approx0.2$~ns as extrapolated from measurements at lower fluences assuming a fluence-proportional trapping probability \cite{Lan10}. Although trapping was recently observed to be lower than expected at high voltages or electric fields, it can be assumed that only very few electrons will reach the back side for all thicknesses studied here.} ($l_{eff}\approx20$~$\upmu$m)$<d$ for all thicknesses $d$ studied, the CCE is given by $CCE=M\cdot l_{eff}/d$. Considering the thickness ratio only, one would expect CCE(75~$\upmu$m)/CCE(100~$\upmu$m)=1.33 and CCE(150~$\upmu$m)/CCE(100~$\upmu$m)=0.66. However, Fig.~\ref{different_material}c shows that e.g. at 600~V the measured CCE ratios are 1.7 and 0.37, respectively, and the difference between the thickness ratio and the actually measured CCE ratio increases with voltage. This suggests that $M$ is higher for thinner diodes. However, some care is needed to explain this effect. The simple expectation that electric fields are higher in thinner sensors at the same applied voltage is only valid if the field extends over the whole sensor thickness. Below full depletion the electric field distribution $E(x)$ in the linear model does not depend on the thickness, but only on the space charge density and none of the extrapolated depletion voltages of the diodes studied here is below 600~V. But modifications to the linear field like double peak and voltage drop over the neutral bulk region are expected to lead to non-zero field values over the entire diode thickness far below the nominal depletion voltage. Moreover, in the case of 75~$\upmu$m another reason for higher maximum fields could be the higher donor introduction rate $g_C$ than in thicker materials, which is probably due to the larger average oxygen concentration. However, in the case of 100 and 150~$\upmu$m diodes the $g_C$ values were found to be very similar.

\section{Annealing behaviour}
\label{Sec:annealing}

\begin{figure}[bt]
	\centering
		\includegraphics[width=8.5cm]{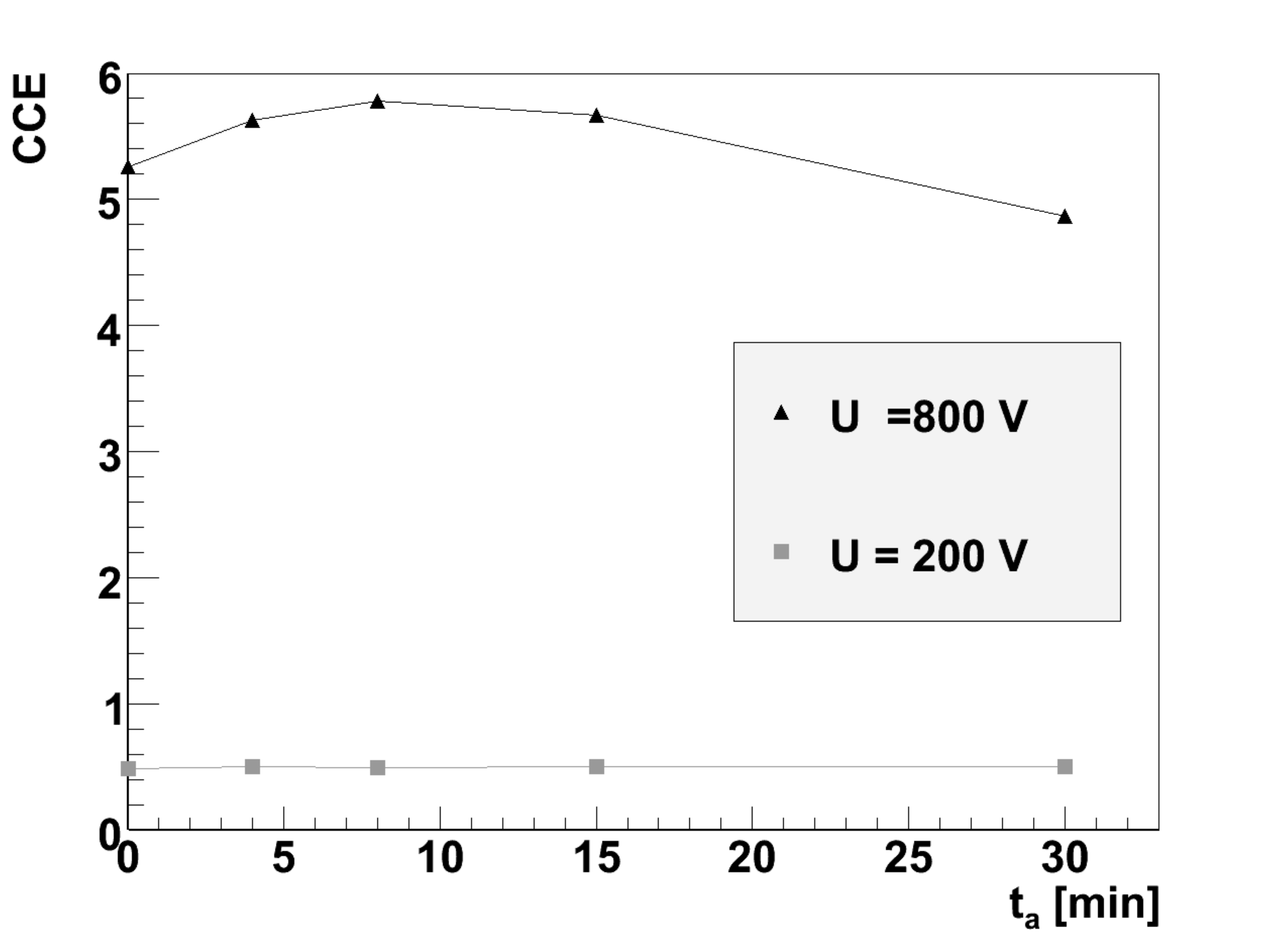}
	\caption{The CCE at 200 and 800~V as a function of annealing time $t_{a}$ for EPI-ST~75~$\upmu$m (small) at $10^{16}$~cm$^{-2}$. Annealing was performed at 80$^\circ$C and CCE measured with 670~nm laser light.}
	\label{annealing}
\end{figure}

The annealing behaviour at 80$^\circ$C of CCE in the CM regime was studied using 670~nm laser light for EPI-ST diodes of 75~$\upmu$m and 100~$\upmu$m thickness and fluences of 6 or $7\times10^{15}$~cm$^{-2}$ and $10^{16}$~cm$^{-2}$. They all show such a characteristic behaviour as displayed in Fig.~\ref{annealing} for the example of EPI-ST~75~$\upmu$m at $10^{16}$~cm$^{-2}$. In the CM regime, e.g. at 800~V, the CCE increases from its value as irradiated up to a maximum at 8~min before it decreases again. This is exactly the same behaviour as observed for the annealing of $U_{dep}$ and the space charge density \cite{Lan08}, at least at lower fluences of up to $4\times10^{15}$~cm$^{-2}$ where CV measurements at room temperature were still possible. This shows again that CM and its underlying high electric fields are closely connected to the space charge which is built up during irradiation and altered during annealing. The annealing behaviour at lower voltages (e.g. 200~V), where CCE is trapping- instead of CM-dominated, is completely different and the CCE does hardly change.

\section{Temperature dependence}
\label{Sec:temperature}

\begin{figure}[bt]
	\centering
		\includegraphics[width=8.5cm]{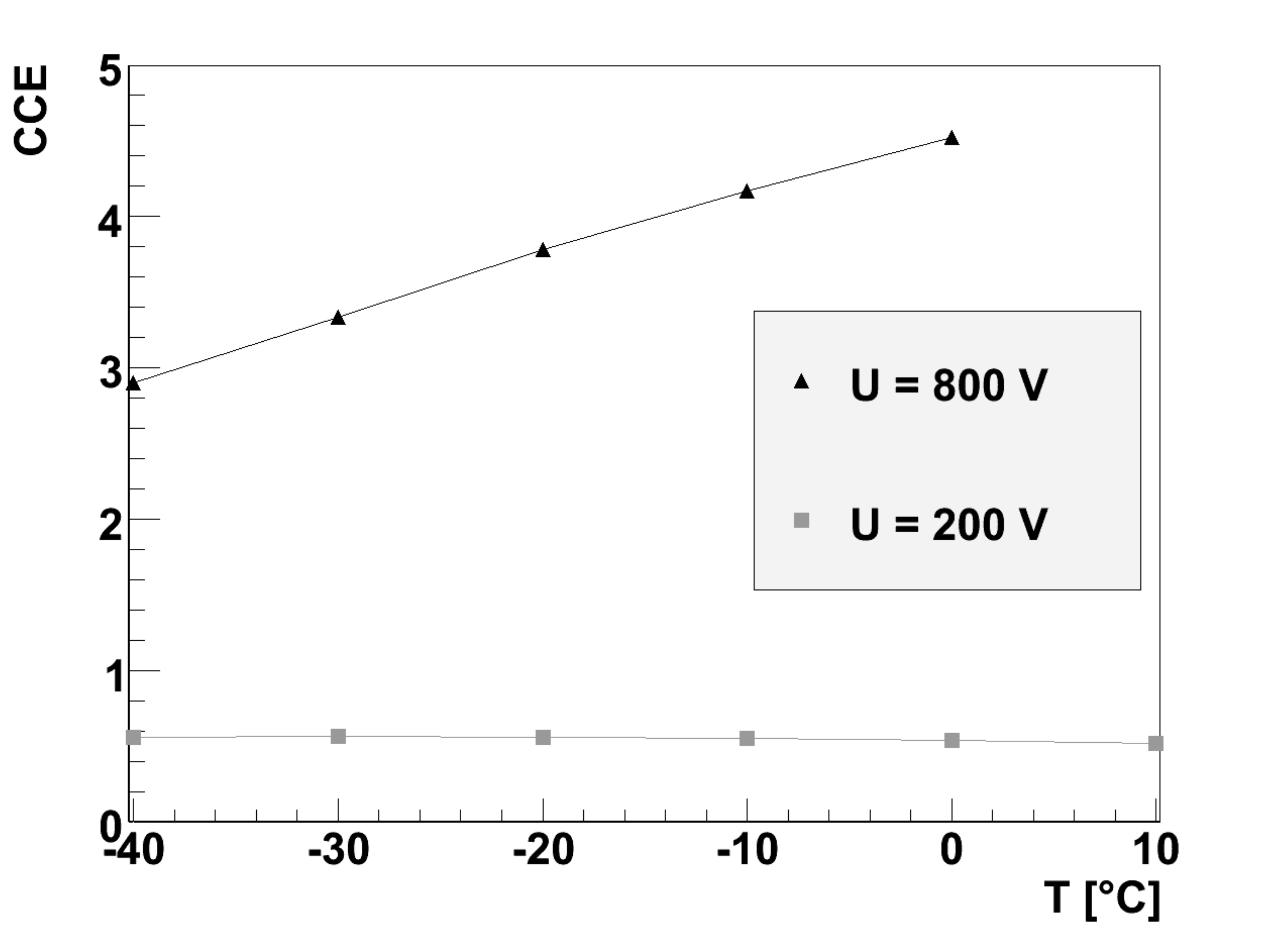}
	\caption{The CCE at 200 and 800~V as a function of temperature for 75-1E16 measured with 670~nm laser light.}
	\label{temperature}
\end{figure}

Fig.~\ref{temperature} shows the temperature dependence of the CCE below and in the CM regime measured with 670~nm laser light. At low voltages (e.g. 200~V) where trapping dominates the charge collection no large sensitivity on temperature can be observed, similarly to the temperature behaviour of CCE at lower fluences \cite{Lan08}. In contrast, at voltages above 300~V a systematic decrease of CCE for decreasing temperature was found. Naively one would expect the opposite behaviour as the ionisation coefficients increase for decreasing temperature \cite{Gra73}. However, on the one hand the absorption length of 670~nm laser light increases slightly for decreasing temperature (according to \cite{Raj79} from 3.1~$\upmu$m at 0$^\circ$C to 3.6~$\upmu$m at -40$^\circ$C) so that at lower temperatures a smaller fraction of charge is produced in the region with the highest electric field at the very front side. Indeed, the effect is less pronounced for 1060~nm laser light and $\alpha$-particles, for which the charge deposition distribution is temperature-independent. However, the trend is the same and a simulation based on the simple model from Sec.~\ref{Sec:Loc_d} suggests that the temperature-dependent change in the absorption length is only a minor correction and cannot compensate for the change in the ionisation coefficient. On the other hand, the electric field could change with temperature. E.g. as already mentioned in Sec.~\ref{Sec:Loc_d}, the leakage current, which is highly temperature-dependent, influences the space charge and thereby the electric field at high fluences.

\section{Linearity of multiplication}
\label{Sec:proportional_mode}

\begin{figure}[bt]
	\centering
		\includegraphics[width=8.5cm]{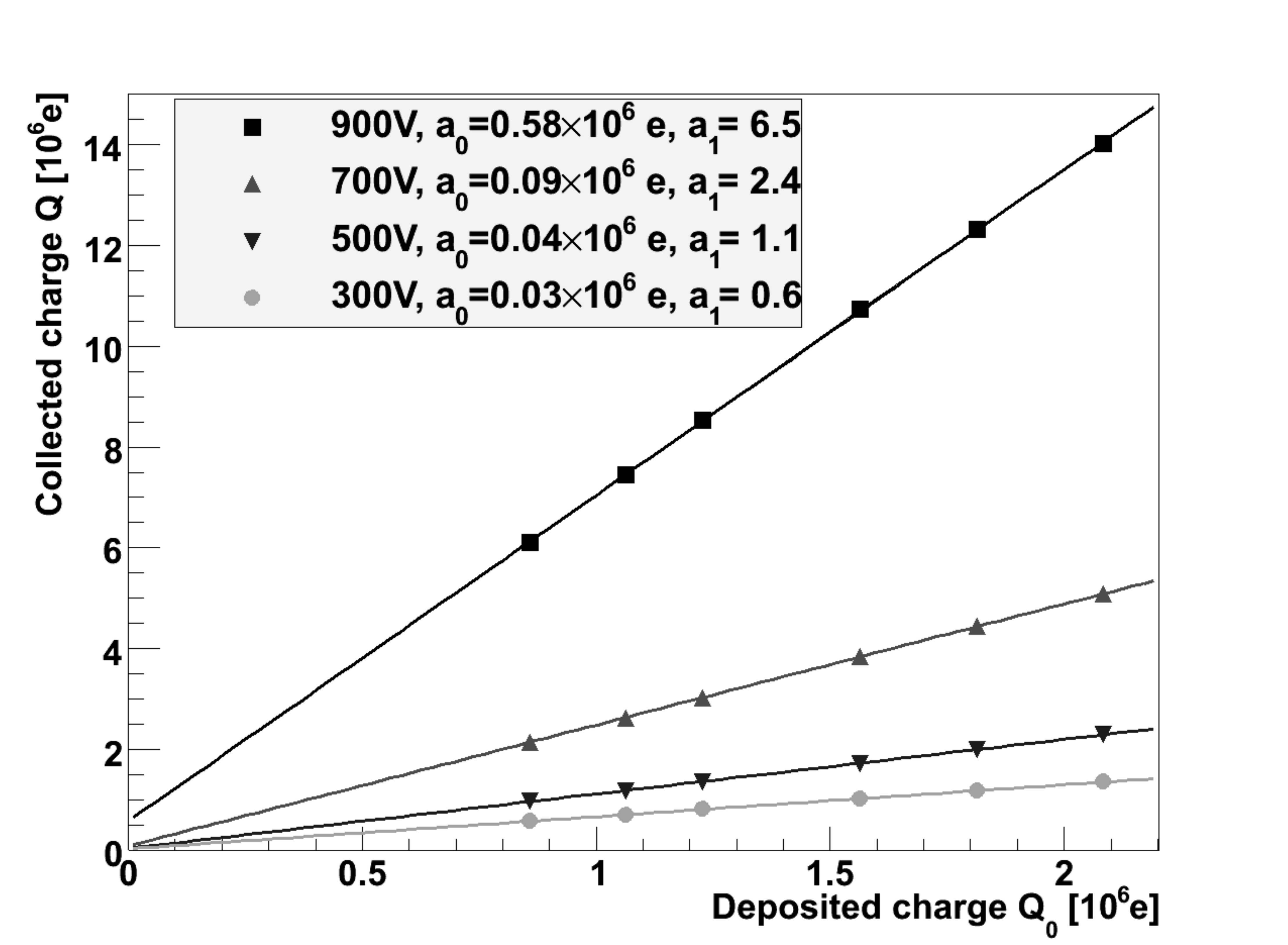}
	\caption{The collected charge as a function of deposited charge with linear fits $Q = a_0 + a_1 \times Q_0$ at different voltages. Measured with 670~nm laser light in 75-1E16.}
	\label{linearity}
\end{figure}

To decide whether a highly-irradiated diode in the CM regime operates in proportional or in Geiger mode, the measured charge $Q$ was plotted as a function of deposited charge $Q_0$ in the range from (0.8 to 2.1)$\times 10^6$~e for 300, 500, 700 and 900~V (see Fig.~\ref{linearity}). 670~nm laser light was used. A linear behaviour can be observed. Although especially the curve at 900~V does not extrapolate exactly to zero, maybe due to noise, the behaviour is approximately proportional. The linear slope corresponds within 10\% to the CCE results in Sec.~\ref{Sec:Loc_d} at the respective voltage. For Geiger mode a substantial multiplication contribution of holes and therefore higher electric fields with a higher $k$-value (i.e. $\alpha_p / \alpha_n$) are needed.

\section{Spatial uniformity over the diode area}
\label{Sec:uniformity}
\begin{figure*}[bt]
	\centering
		a)\includegraphics[width=8.5cm]{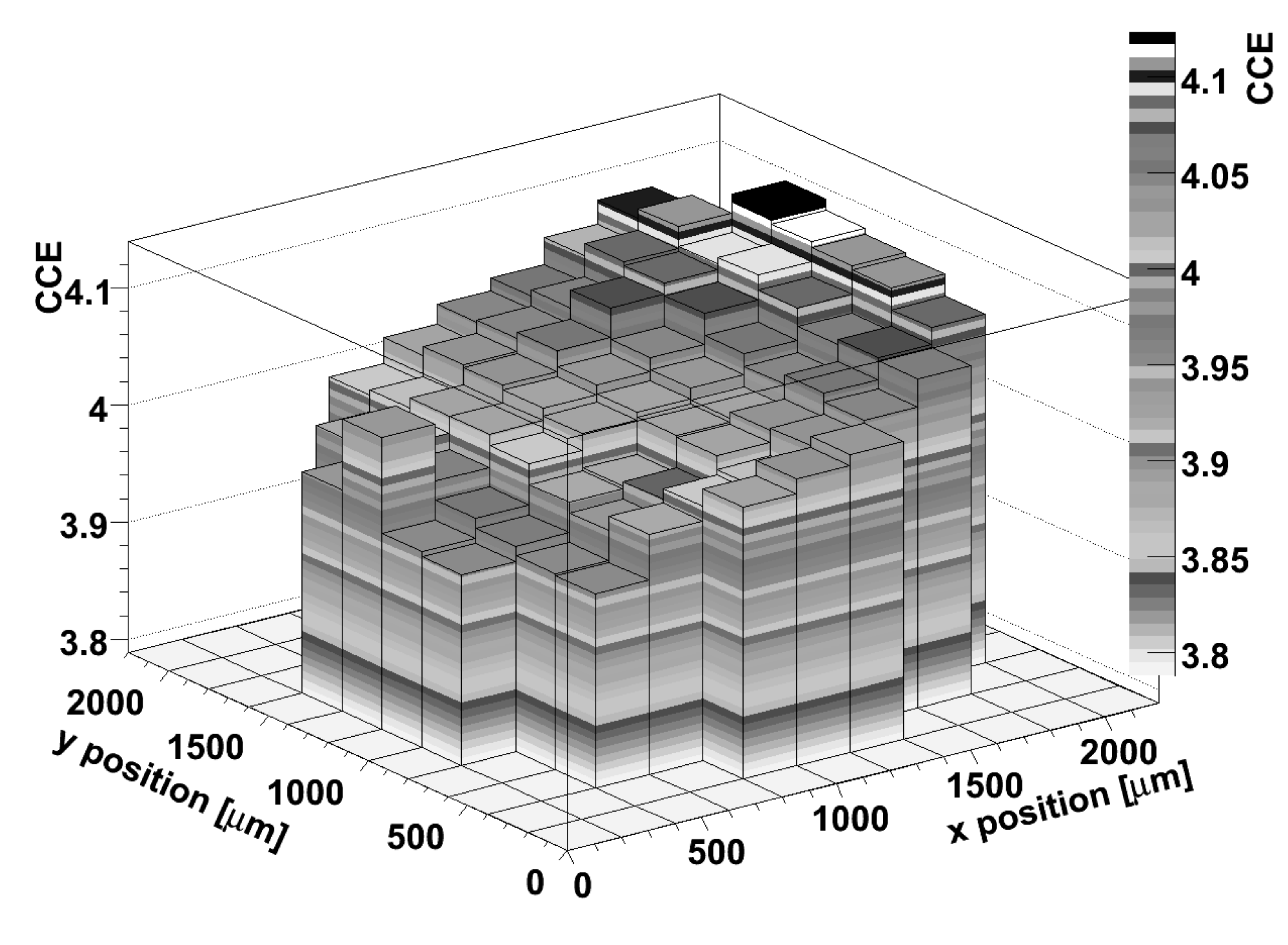}
		b)\includegraphics[width=8.5cm]{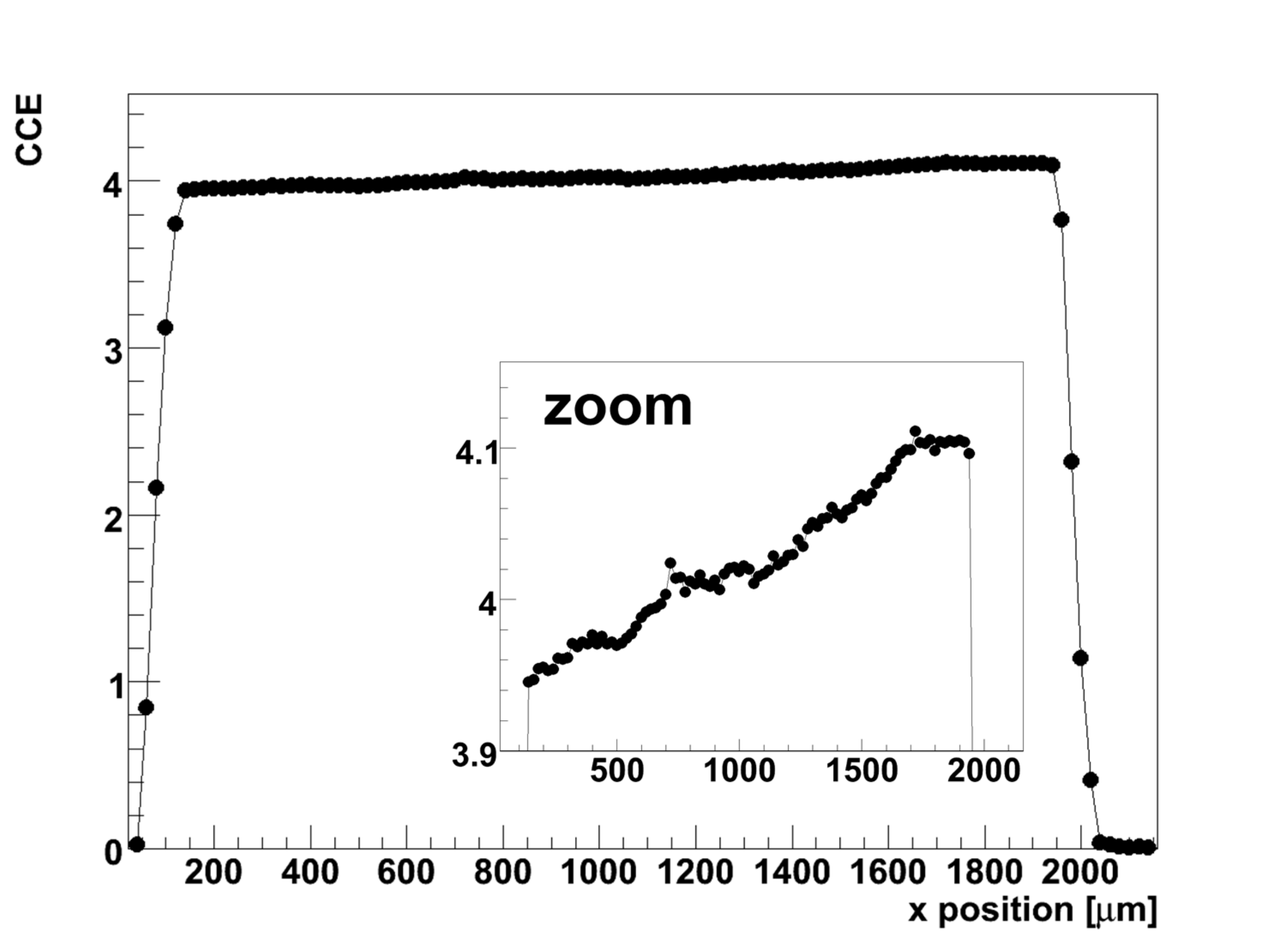}
	\caption{a) X-y scan over the illumination window with a focused laser spot of 660~nm wavelength at 800~V in 75-1E16. b) X-scan over the middle of the window at the same voltage in the same diode.}
	\label{x-y-scan}
\end{figure*}

In order to investigate the spatial uniformity of CM over the diode area, an x-y scan was performed over the illumination window in the front metallisation of the diode (75-1E16) with a focused 660~nm laser (spot size $\sigma_{spot}$=20~$\upmu$m) and a step width of 200~$\upmu$m at 480, 580, 650 and 800~V corresponding to mean CCE values of 1.0, 1.5, 2.2 and 4.0. Fig.~\ref{x-y-scan}a shows for the example of 800~V that the CCE is very uniform over the whole area (note the zoom in the CCE scale). The normalised standard deviation $\sigma/\overline{CCE}$ is around 0.5\% for 480, 580 and 650~V and 1\% for 800~V. However, a closer look reveals that there seems to be a systematic, almost linear slope in x-direction as demonstrated by an x-scan with a much finer step width of 10~$\upmu$m (Fig.~\ref{x-y-scan}b). But the normalised slope is rather small with only 0.5\% difference over 1~mm in the middle of the window for 480~V, increasing up to 1.7\% difference over 1~mm for 800~V. A reason for this behaviour might be nonuniform irradiation.

\section{Long-term stability}
\label{Sec:Stability}

\begin{figure*}[bt]
	\centering
		a)\includegraphics[width=8.5cm]{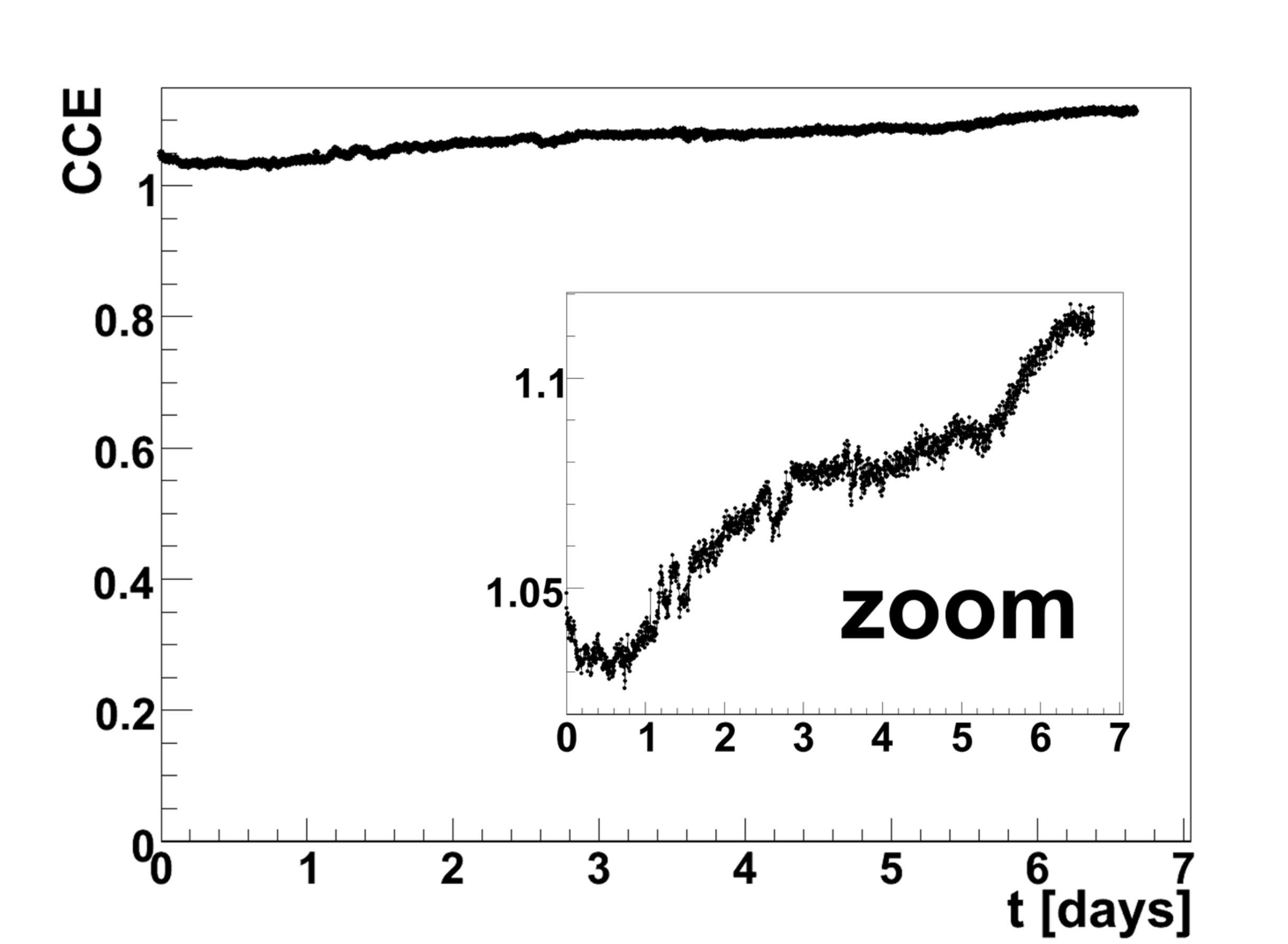}
		b)\includegraphics[width=8.5cm]{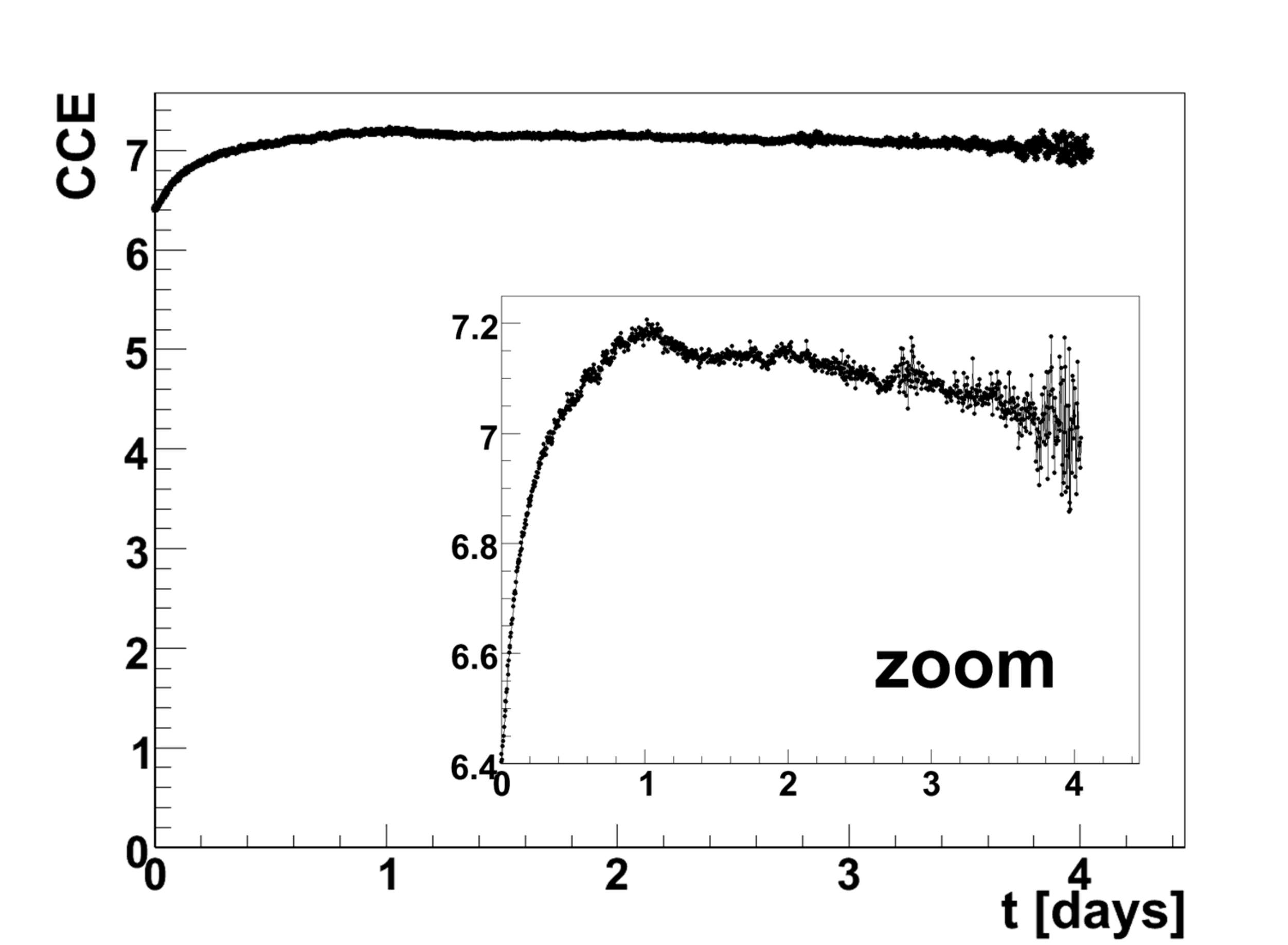}
	\caption{The stability of collected charge measured with 670~nm laser light over several days in 75-1E16 at 480~V (a) and 900~V (b).}
	\label{stability}
\end{figure*}

For an application of the CM effect in a real detector it is important that it remains reliably stable over a long period of time. Fig.~\ref{stability} shows a long-term measurement over several days in 75-1E16 with 670~nm laser light. Every 5~min a measurement was taken. The diode was kept at a constant bias voltage of 480~V which corresponds to CCE$\approx$1 (Fig.~\ref{stability}a) or 900~V corresponding to CCE$\approx$7 (Fig.~\ref{stability}b). The temperature around -11$^\circ$C was stable within 0.3$^\circ$C and the leakage current within 3\% (480~V) or 10\% (900~V).

It can be seen that the CCE at 480~V drops during the first half day from 1.05 to 1.03 and then increases more or less monotonically about 8\% up to 1.11. However, over such a long period of time the laser intensity might change. Reference measurements were performed with an unirradiated diode, for which the collected charge after the long-term measurement with the irradiated sample was also about 7\% higher than before. To conclude, the operation in the CM regime at 480~V can be regarded as stable.

At 900~V there is a rise of about 12\% from 6.4 to 7.2 during the first day before a decrease to 7.0 after 4 days. However, at such a high voltage micro discharges occurred (see Sec.~\ref{Sec:methods}). At the beginning their rate was low enough not to be visible in the externally triggered measurements. But after about 2.5 days their rate became so high that they often interfered with a TCT pulse. As 512 externally triggered pulses have been averaged, this becomes manifest only as moderate, but clearly visible fluctuations of the CCE curve. However, in the case of self-triggering or high-frequency readout like at the (S)LHC it would be impossible to detect the signal of an impinging particle reliably and it poses potential dangers for the readout electronics. For operation at such high voltages, micro discharges would have to be avoided.

\section{Charge spectrum and noise}
\label{Sec:charge-spectrum-noise}
So far only the signal averaged over 512 pulses was discussed. However, in a real experiment a detector signal needs to be separated from noise event-by-event. Therefore, 301 single TCT pulses were taken and their baseline noise and charge spectrum investigated with special emphasis on the CM regime. 

\subsection{Baseline noise}
\label{Sec:Noise}

\begin{figure*}[bt]
	\centering
		a)\includegraphics[width=8.5cm]{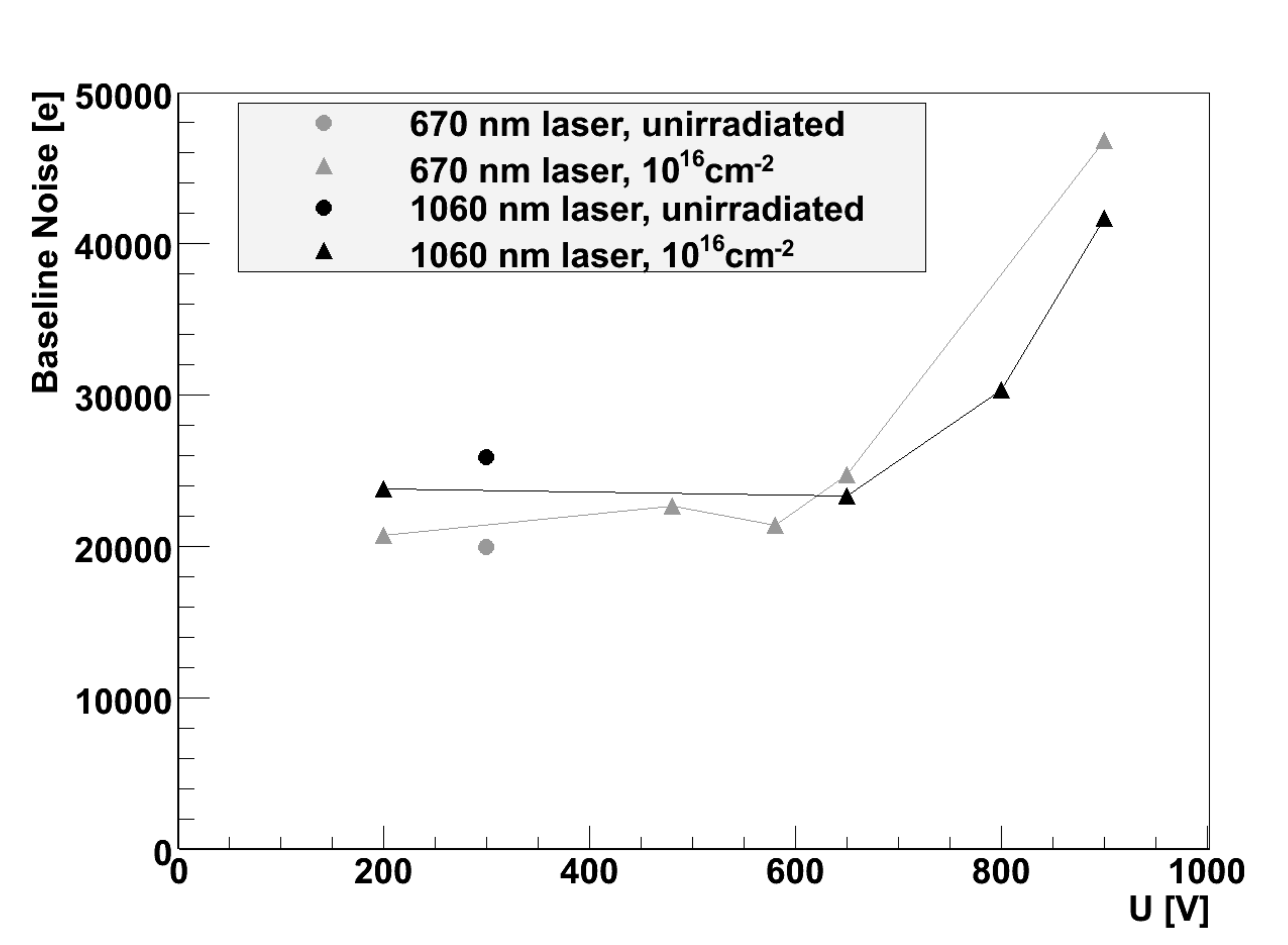}
		b)\includegraphics[width=8.5cm]{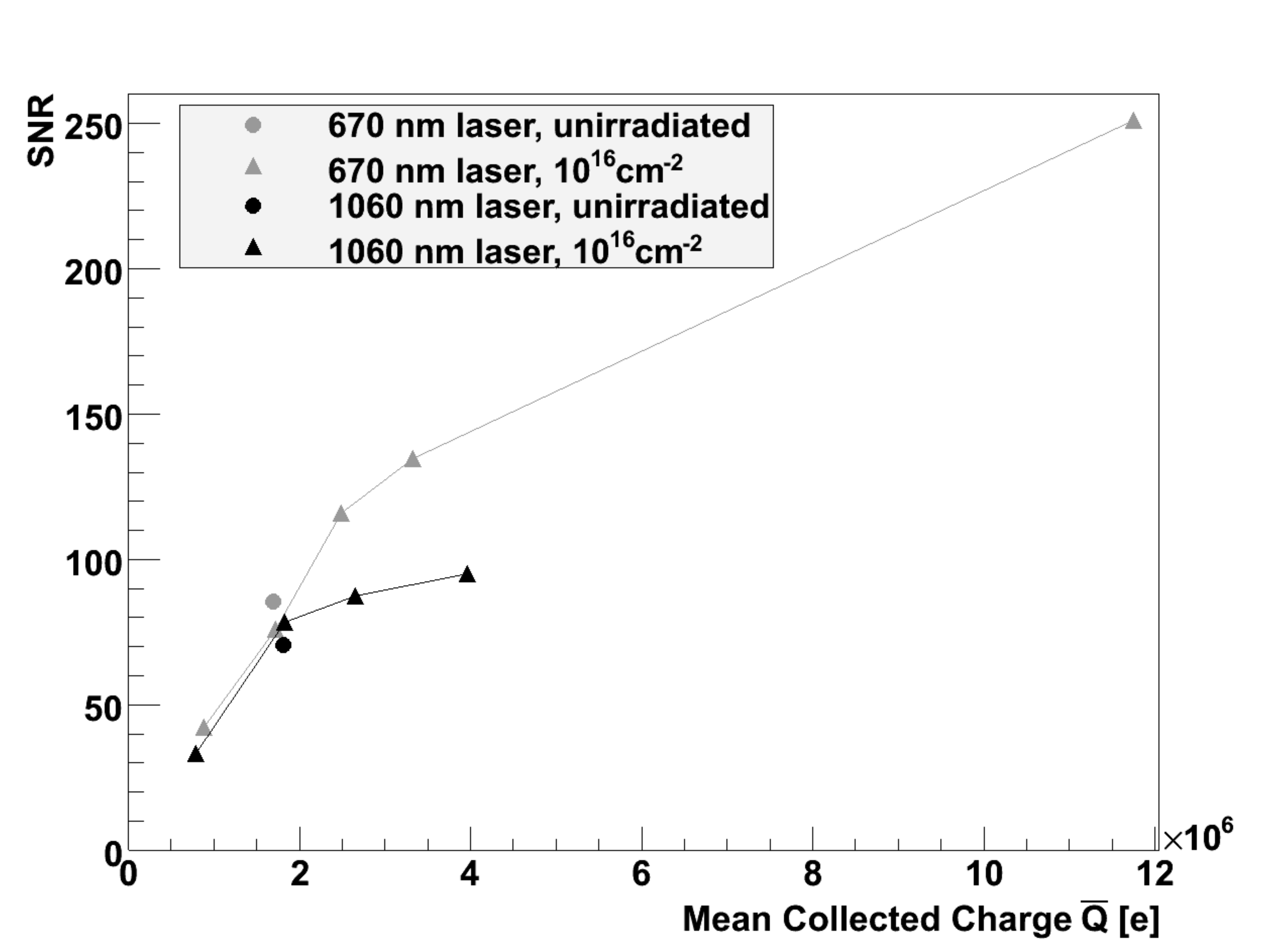}
	\caption{a) The RMS noise of the TCT baseline in n-EPI-ST~75~$\upmu$m (unirradiated and 75-1E16) as a function of voltage. b) The corresponding SNR as a function of the mean collected charge for $\approx 1.8\times 10^6$~e-h pairs deposited.}
	\label{noise}
\end{figure*}

\begin{figure*}[tb]
	\centering
		a)\includegraphics[width=8.5cm]{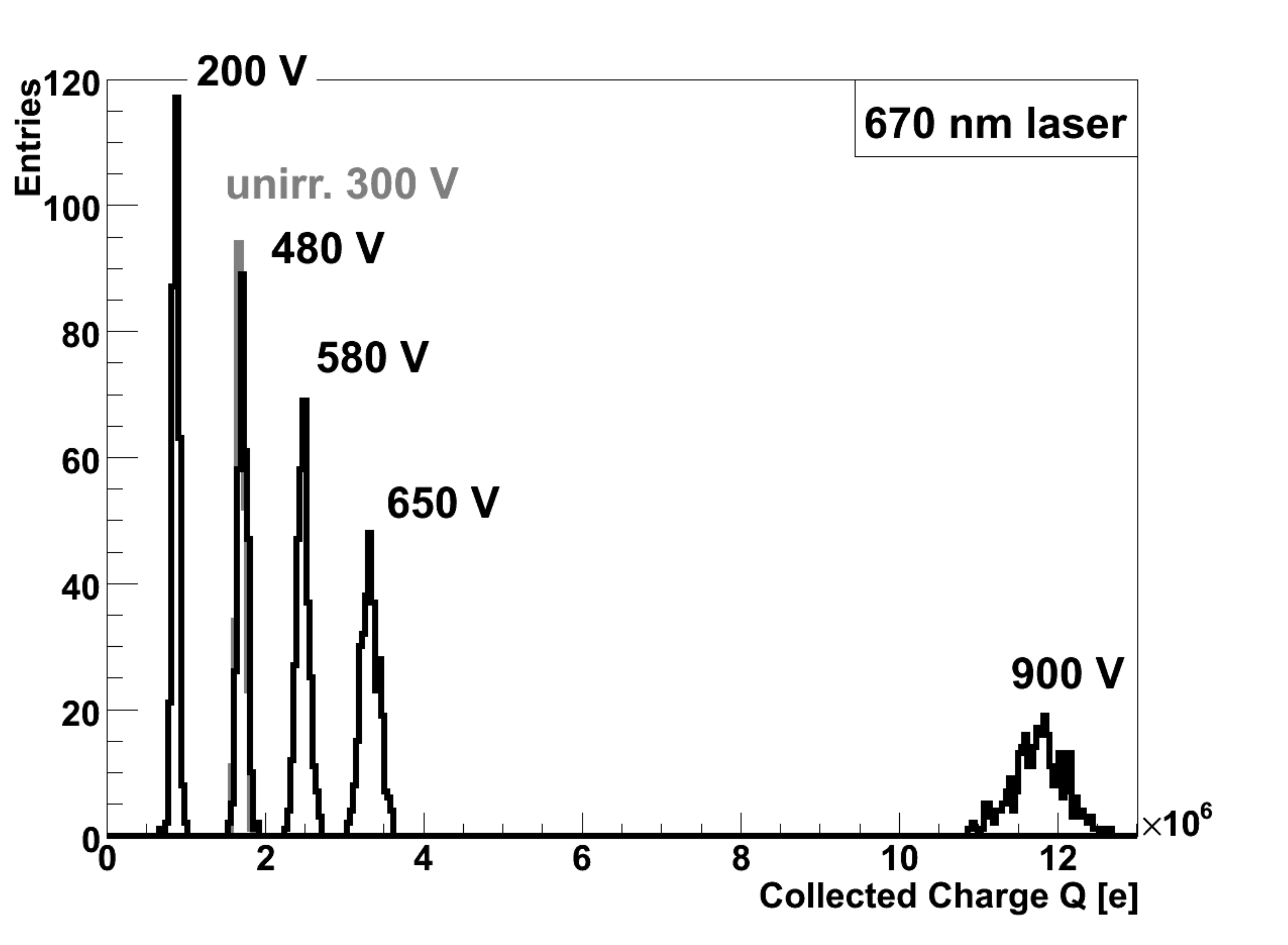}
		b)\includegraphics[width=8.5cm]{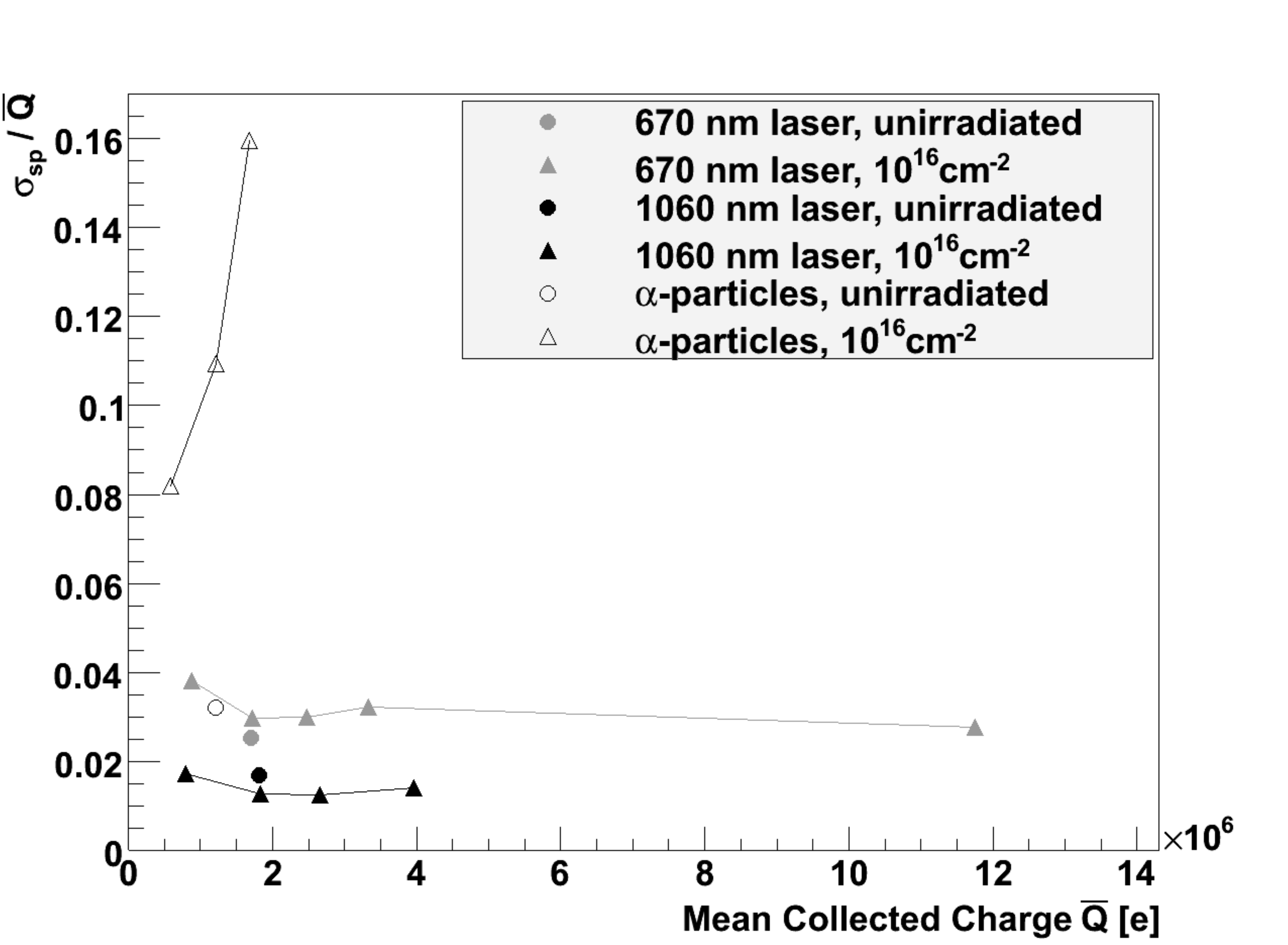}
	\caption{a) The charge spectra measured with 670~nm laser light in 75-1E16 at different voltages and in an unirradiated 75~$\upmu$m diode at 300~V, which almost entirely overlaps with the spectrum of 75-1E16 at 480~V. b) The relative width $\sigma_{sp}/\bar{Q}$ (corrected for noise) of the charge spectra as a function of the mean collected charge $\bar{Q}$ for 670 and 1060~nm laser light and $\alpha$-particles. In the case of $\alpha$-particles single events with lower charge than in the central peak were cut off in order to exclude low-energy particles and micro discharges.}
	\label{spectra}
\end{figure*}

The shot noise\footnote{Noise is assumed to be in units of charge (\emph{equivalent noise charge, ENC}) in the following.} $\sigma_{sh}$ related to fluctuations in the bulk leakage current is expected to increase in the CM regime because also the bulk leakage current is multiplied (with a factor $M^\prime$ which can be different from $M$ for the signal). Already a constant multiplication would increase the shot noise by $M^\prime$, which might be further enhanced by fluctuations due to the statistical nature of CM, which are described by the excess noise factor~$F^\prime$ \cite{Web74}:

\begin{equation}
\sigma_{sh}(M^{\prime})=\sigma_{sh,M^{\prime}=1}M^\prime\sqrt{F^\prime} ,
\end{equation}
where $\sigma_{sh,M^{\prime}=1}\propto \sqrt{I_{rev,M^{\prime}=1}}$ is the shot noise without CM. According to \cite{McI66} $F^\prime$ is increasing with $M^{\prime}$ and $k$. For $k \approx 0$, which is valid for low fields just above the CM threshold, $F^\prime$ goes asymptotically to 2 for large $M^{\prime}$. However, $k$ is a strong function of $E$ so that also $F^\prime$ is increasing with increasing field.

The signal-to-noise ratio is given by

\begin{equation}
SNR = \frac{\bar{Q}}{\sigma_{noise}} = \frac{M\bar{Q}_{M=1}}{\sqrt{\sigma_{sh,M^{\prime}=1}^2M^{\prime2}F^\prime + \sigma_{noise}^{\prime2}}} ,
\end{equation}
where $\bar{Q}$ is the mean collected charge, $\bar{Q}_{M=1}$ the mean collected charge without CM and $\sigma_{noise}^{\prime}$ noise contributions that are not affected by CM. Thus, it depends on the relation of the parameter values whether CM can improve the SNR. For $M\approx M^{\prime}$ the SNR improves as long as $\sigma_{sh}(M^{\prime})<\sigma_{noise}^{\prime}$. If $\sigma_{sh}(M^{\prime}) \gg \sigma_{noise}^{\prime}$ the SNR can only improve if $M>M^{\prime}\sqrt{F^\prime}$.

With a TCT setup the transient current pulse is recorded directly. Thus, the ENC was determined in a similar way as the charge signal by integrating over the TCT baseline before the pulse in a time window of $\approx$10~ns. Fig.~\ref{noise}a shows the RMS noise as a function of bias voltage for measurements with 670 and 1060~nm laser light. The difference is probably due to random fluctuations. The noise of an unirradiated diode, which should be mostly dominated by other noise contributions $\sigma_{noise}^{\prime}$ than shot noise since leakage current is low, was found to be within 20-26~ke. This shows that the noise of a TCT setup is significantly higher than in the case of a charge-sensitive readout with optimised shaper. Up to 650~V the noise of 75-1E16 is similar to the one of the unirradiated diode before increasing to $\approx$45~ke at 900~V. This behaviour can be explained by the high value of $\sigma_{noise}^{\prime}$ of the TCT setup, which is exceeded by an increasing shot noise only above 650~V.

From this, the SNR was calculated, which is shown in Fig.~\ref{noise}b as a function of the mean collected charge $\bar{Q}$. As for both 670 and 1060~nm laser light similar values of $Q_0\approx1.8\times10^{6}$~e-h pairs were deposited, the SNR for the unirradiated diodes are also similar (70-85). Even for 75-1E16 similar SNR values were found at the same collected charge of $Q_0$ because at the voltages needed for CCE values of 1 (480~V for 670~nm and 650~V for 1060~nm) the noise has not increased. For higher voltages respectively mean collected charge the SNR is still improving, but slower than before because also the noise is increasing. For the same collected charge, the SNR of 670~nm is seen to be higher than the SNR of 1060~nm because in the case of 670~nm lower voltages are needed for the same CCE (cf. Fig.~\ref{CCE_different_pen}), which implies lower noise. But it should be emphasised that also in the case of 1060~nm laser light, which has similar penetration as MIPs, in highly-irradiated sensors at high voltages higher SNR values are obtained than before irradiation due to CM. However, for the TCT setup with laser both the noise and the signal were much higher than in a realistic SLHC detector so that further studies with MIPs and a low-noise charge readout will be carried out.

\subsection{Charge spectra}
\label{Sec:charge-spectrum}

CM does not only affect the electronic noise of the baseline, but also the charge spectrum of the signal. The width of the spectrum is expected to increase due to multiplication and associated statistical fluctuations in a similar way as the shot noise (but note the different $M$ and $F$): $\sigma_{sp}=\sigma_{sp,M=1}M\sqrt{F}$. Furthermore, when performing a measurement the charge spectrum is convoluted with the baseline noise so that the measured width is given by $\sigma_{sp,meas}=\sqrt{\sigma_{sp}^2 + \sigma_{noise}^2}$.

Fig.~\ref{spectra}a shows the charge spectra measured with 670~nm laser light for an unirradiated diode at 300~V and for 75-1E16 at different voltages. The charge spectrum width (standard deviation) for the unirradiated diode is 48~ke. The broadening due to CM is easily visible when comparing the spectra of 75-1E16 at low and high voltages. To investigate whether the broadening is enhanced by fluctuations, the relative width $\sigma_{sp} / \bar{Q}$ was calculated (after correcting for baseline noise) because both $\sigma_{sp}$ and $\bar{Q}$ are proportional to $M$ so that $M$ cancels. The relative width is shown in Fig.~\ref{spectra}b as a function of mean collected charge for 670 and 1060~nm laser light and $\alpha$-particles. For laser light it can be seen that at $Q_0\approx1.8\times10^{6}$~e (CCE=1) the relative width of 75-1E16 is not much higher than before irradiation (670~nm) or even slightly lower (1060~nm). The differences might be due to measurement uncertainties. Moreover, the relative width of 75-1E16 stays constant even for higher $\bar{Q}$ (i.e. higher voltages and multiplication), which shows that in the range considered here CM fluctuations are not dominant, probably due to low values of $M$ and $k$ at not too high fields.

In contrast, the relative width of the spectrum measured with $\alpha$-particles significantly increases in the CM regime. Statistical fluctuations in the CM process were ruled out by the measurements with laser light. Thus, the large broadening is probably caused by fluctuations in the amount of charge deposited in the thin CM region. These might be caused by the divergence of the beam which leads to varying penetration of the $\alpha$-particles (see Sec.~\ref{Sec:methods}). E.g. the bragg peak depth is reduced from 26~$\upmu$m at perpendicular incidence to 18~$\upmu$m at 45$^\circ$, which makes a significant difference as the CM region is assumed to be in a similar thickness range and becomes increasingly stronger for decreasing depth.

It needs to be studied further how the charge spectrum in the CM regime looks for MIPs as in this case Landau fluctuations might lead to a varying deposition in the CM region as well, especially as the layer is so thin. Moreover, the charge deposited by MIPs is much lower (most probable value$\approx$80~e/$\upmu$m) so that much smaller variations of noise and charge spectrum width than measurable here might play a role.

\section{Conclusions}
Charge multiplication (CM) in thin epitaxial n-type pad diodes irradiated by protons up to $10^{16}$~cm$^{-2}$ equivalent fluence was investigated by CCE measurements with different TCT setups. Using radiation of different penetration like 670, 830 and 1060~nm laser light and $\alpha$-particles with and without absorbers, the CM region could be localised near the p$^+$-implant. CM was observed to be higher for thinner sensors and for DO material compared to ST material and the annealing curve at 80$^\circ$C showed the same behaviour as the one for $U_{dep}$. All this confirms the assumption of CM as a combined effect of thin sensors and large radiation-induced space charge densities, leading to high electric fields. For a deeper understanding of the formation of a radiation-induced CM region a better knowledge of the electric field is required, which takes effects like double peak and voltage drop over the neutral bulk region into account. Conclusive hints could come from the recently developed method of Edge-TCT \cite{Kra09}. The CCE in the CM regime measured with 670~nm laser light was found to decrease at lower temperatures. The collected charge was found to be proportional to the deposited charge. Using an x-y scan, the collected charge in the CM regime was found to be uniform with deviations on the percent level. Long-term measurements showed that the CCE remained stable over many days. The biggest challenge for the operation of detectors in the CM regime were micro discharges which occurred randomly at high voltages. Improvements of the device technology to avoid them would be desirable. The noise of the TCT baseline was found to increase above 650~V, but the SNR was seen to improve due to CM up to 900~V for 670 and 1060~nm laser light. The relative width of charge spectra in the CM regime remained constant in the case of laser light, which shows that statistical fluctuations in the CM process are negligible for injected charge in the order of $10^{6}$~e. However, in the case of $\alpha$-particles the relative width increased, probably due to varying charge deposited in the CM region.

To conclude, CM emerging in highly-irradiated detectors seems to be a promising option to overcome trapping at SLHC fluences. Very good detector performances were demonstrated with laser light. Measurements with $\beta$-particles are in progress in order to study the collected charge in the CM regime for MIP-like particles and the noise with a low-noise charge readout. Effects of CM on the position resolution will need to be investigated with highly-irradiated segmented detectors.

\section*{Acknowledgements}
We would like to thank M. Glaser for his help with the proton irradiations. Partial funding by the German Ministry of Education and Research (BMBF) under the project "FSP 102 - CMS Detektor am LHC", by the HGF Alliance "Physics at the Terascale", by the CiS Hamburg project under Contract no. SSD 0517/03/05 and by the European XFEL is gratefully acknowledged.


\begin{thebibliography}{00}

\bibitem{Gia02}
F. Gianotti et al., hep-ph/02004087, April 2002.

\bibitem{Lan08}
J. Lange, Radiation Damage in Proton-Irradiated Epitaxial Silicon Detectors, Diploma thesis, University of Hamburg, October 2008, DESY-THESIS-2009-022.

\bibitem{Lan10}
J. Lange et al., Nucl. Instr. and Meth. A (2010), doi:10.1016/j.nima.2009.11.082.

\bibitem{Man09}
I. Mandi\'{c} et al., Nucl. Instr. and Meth. A 603 (2009) 263.

\bibitem{Kra09}
G. Kramberger et al., IEEE NSS Conference Record, N25-206, Orlando, USA, 2009.

\bibitem{Mik10}
M. Miku$\check{\mbox{z}}$ et al., Nucl. Instr. and Meth. A (2010), doi:10.1016/j.nima.2010.04.084.

\bibitem{Cas09}
G. Casse et al., Nucl. Instr. and Meth. A (2010), doi:10.1016/j.nima.2010.02.134.

\bibitem{Koe10}
M. K\"{o}hler et al., Test Beam and Laser Measurements With Irradiated 3D Silicon Strip Detectors, presented at the 16th RD50 Workshop, Barcelona, June 2010, http://indico.cern.ch/getFile.py/access?con tribId=30\&sessionId=6\&resId=0\&materialId=slides\&confId=86625.

\bibitem{Lin06}
G. Lindstr\"{o}m et al., Nucl. Instr. and Meth. A 568 (2006) 66.

\bibitem{Pin09}
I. Pintilie et al., Nucl. Instr. and Meth. A 611 (2009) 52.

\bibitem{ITME}
Institute of Electronic Materials Technology (ITME), 133 Wolczynska Str., 01-919 Warsaw, Poland.

\bibitem{CiS}
Forschungsinstitut f\"{u}r Mikrosensorik und Photovoltaik GmbH (CiS), Konrad-Zuse-Str. 14, 99099 Erfurt, Germany.

\bibitem{PIrS}
M. Glaser. http://irradiation.web.cern.ch/irradiation.

\bibitem{Ram39}
S. Ramo, Proc. IRE 27 (1939) 584.

\bibitem{Kra02}
G. Kramberger et al., Nucl. Instr. and Meth. A 481 (2002) 297.

\bibitem{Gra73}
W. N. Grant, Solid-St. Electron. 16 (1973) 1189.

\bibitem{SRIM}
J. F. Ziegler, SRIM-2008. The Stopping and Range of Ions in Matter, 2008. www.srim.org.

\bibitem{Ere02}
V. Eremin et al., Nucl. Instr. and Meth. A 476 (2002) 556.

\bibitem{Ver07}
E. Verbitskaya et al., Nucl. Instr. and Meth. A 583 (2007) 77.

\bibitem{Raj79}
K. Rajkanan et al., Solid-St. Electron. 22 (1979) 793.

\bibitem{Web74}
P. P. Webb et al., RCA Review 35 (1974) 234.

\bibitem{McI66}
R. J. McIntyre, IEEE Trans. Electron Dev. ED-13 (1966) 164.



\end{thebibliography}
\end{document}